\documentclass[11pt, oneside]{article}   	

\usepackage{jheppub}
\usepackage{graphicx}				

\usepackage{hyperref}

\usepackage{latexsym, amsmath, amssymb, mathrsfs}
\usepackage[utf8]{inputenc}
\newcommand{\beq}{\begin{eqnarray}}
\newcommand{\eeq}{\end{eqnarray}}
\usepackage{xcolor}
\usepackage{esint}
\usepackage{empheq}
\usepackage{subcaption}
\usepackage[normalem]{ulem}
\usepackage{comment}

\newcommand{\res}[1]{\underset{#1}{\rm res}}

\newcommand{\bfa}{\boldsymbol{\alpha}}
\newcommand{\bfg}{\boldsymbol{\gamma}}

\newcommand{\bft}{\boldsymbol{\theta}}

\newcommand{\ordered}[1]{\{\overrightarrow{#1}\}}
\newcommand{\orderedpi}[1]{\{\overrightarrow{#1} + i\pi\}}
\newcommand{\pordered}[1]{\{\overleftarrow{#1}\}}
\newcommand{\porderedpi}[1]{\{\overleftarrow{#1} + i\pi\}}

\newcommand{\stkout}[1]{\ifmmode\text{\sout{\ensuremath{#1}}}\else\sout{#1}\fi}
\newcommand{\twt}{\tilde\theta}

\title{Extending the Thermodynamic Form Factor Bootstrap Program: Multiple particle-hole excitations, crossing symmetry, and reparameterization invariance.}

\author[a]{Mi{\l}osz Panfil}
\author[b]{and Robert M. Konik}

\affiliation[a]{Faculty of Physics, University of Warsaw, ul. Pasteura 5, 02-093 Warsaw, Poland}
\affiliation[b]{Condensed Matter Physics and Materials Science Division, Brookhaven National Laboratory, \\Upton, NY 11973, USA}

\emailAdd{milosz.panfil@fuw.edu.pl}
\emailAdd{rmk@bnl.gov}

\date{\today}							

\abstract{In this study, we further the thermodynamic bootstrap program which involves a set of recently developed ideas used to determine thermodynamic form factors of local operators in integrable quantum field theories. These form factors are essential building blocks for dynamic correlation functions at finite temperatures or non-equilibrium stationary states. In this work we extend this program in three ways. 
Firstly, we demonstrate that the conjectured annihilation pole axiom is valid in the low energy particle-hole excitations.  Secondly, we introduce a crossing relation, which establishes a connection between form factors with different excitation content. Typically, the crossing relation is a consequence of Lorentz invariance, but due to the finite energy density of the considered states, Lorentz invariance is broken. Nonetheless a crossing relation involving excitations with both particles and holes can established using the finite volume representation of the thermodynamic form factors.  Finally, we demonstrate that the thermodynamic form factors satisfy a reparameterization invariance, an invariance which encompasses crossing.  Reparameterization invariance exploits the fact that the details of the representation of the thermodynamic state are unimportant.
In the course of developing these results, we demonstrate the internal consistency of the thermodynamic form factor bootstrap program in a number of ways. Finally, we provide explicit computations of form factors of conserved charges and densities with crossed excitations and show our results can be used to infer information about thermodynamic form factors in the Lieb-Liniger model.}

\begin{document}

\maketitle
\flushbottom

\section{Introduction}

In this work we study $(1+1)d$ quantum field theories (QFT's) at finite temperatures, or more generally, in states of finite energy density. The standard techniques of QFT to address such situations include the Matsubara formalism, thermofield doubles and more recently the AdS/CFT correspondence~\cite{TGT_BOOK,HQM_BOOK}. Formally, these methods apply to many different systems, but in practice they are limited to either perturbative computations or large $N$ expansions. Our strategy here is to instead focus on integrable quantum field theories (IQFT's), whose vacuum physics is well understood even in the strongly interacting regime. In $(1+1)d$ they form a large subset of quantum field theories, and indeed $(1+1)d$ conformal field theories can be viewed as their massless counterparts. 

Integrable quantum field theories exhibit an elastic and factorizable scattering of their asymptotic particles~\cite{SmirnovBOOK,MussardoBOOK}. This feature has two important consequences. Firstly, the form factors of local operators satisfy certain relations known as the bootstrap axioms. These relations are strong enough to determine the form factors between the vacuum and states with a finite number of particles. Secondly, one can study the thermodynamics of IQFTs, which allows for the calculation of the free energy using the thermodynamic Bethe ansatz (TBA). Remarkably, these two approaches can be combined within the LeClair-Mussardo formula~\cite{LECLAIR1999624} to compute thermal thermal expectation values (one-point functions)
\begin{equation} \label{LcM}
    \langle \mathcal{O} \rangle = \sum_{k=0}^{\infty} \int \prod_{j=1}^k \left(\frac{d\theta_j}{2\pi} n(\theta_j)\right) f_c(\theta_1, \dots, \theta_k),
\end{equation}
where $n(\theta)$ is the thermal filling function and $f_c(\theta_1, \dots, \theta_k)$ are the connected form factors of the operator $\mathcal{O}$ between vacuum and asymptotic state with particles of rapidities $\{\theta_1, \dots, \theta_k\}$. The thermal expectation value is then computed, through a spectral sum, by summing contributions from an arbitrary number of fluctuations above the vacuum state. Interestingly enough the sums can be computed to high accuracy by only including the first terms~\cite{MussardoBOOK,PhysRevLett.111.100401}. The LeClair-Mussardo was subsequently rigorously derived~\cite{Pozsgay:2010xd} and generalized to non-equilibrium stationary states~\cite{PhysRevLett.111.100401} and to theories with internal symmetries~\cite{HUTSALYUK2021115306}.  Furthermore, for the vertex operators and their descendants in the sinh-Gordon model, the infinite series can be resummed~\cite{Negro_2013,Negro_2014}.

The LeClair-Mussardo formula is notable because it calculates thermal expectation values using vacuum form factors. However, its straightforward generalization is not applicable to two- and higher-point functions~\cite{Saleur:1999hq}.  Vacuum form factors can be used to compute two-point functions at finite temperature in certain select cases. The zero-frequency limit of the susceptibility in the O(3) non-linear sigma model (NLSM) \cite{PhysRevB.68.104435} at low but finite temperature can be computed with a finite number of vacuum form factors while the finite temperature line shape of the spin-spin correlation function in the quantum Ising and the O(3) NLSM can be computed by resumming an infinite class of vacuum form factors \cite{PhysRevB.75.144403,Essler2009}.  Similarly the asymptotics of transport coefficients in the sine-Gordon model can be computed through such resummations \cite{Altshuler2006}. In a similar spirit the vacuum form-factors can be used to determine the non-equilibrium dynamics following a small quantum quench~\cite{Bertini_2014, Bertini:2016xgd}.

In general such resummation methods are challenging and it is desirable to derive directly form factors which involve excitations not relative to the vacuum state but relative to a finite energy density state.
These modified form factors are known as thermodynamic form factors (TDFFs). 
The two-point function in the spectral representation involving TDFFs takes the following form:
\begin{equation} \label{2point_fnc}
    \langle \mathcal{O}(x,t) \mathcal{O}(0,0) \rangle_c = \sum_{k=1}^{\infty} \sum_{m=1}^{\infty} \int {\rm d} \bft_k^+ {\rm d} \bft_m^- e^{i d(x,t; \bft_k^+, \bft_m^-)}|f_{\rho_{\rm p}}(\bft_k^+, \bft_m^-)|^2,
\end{equation}
where $\bft_k$ denotes a set $\{\theta_1, \dots \theta_k\}$ of cardinality $k$ with $\bft_k^+$ particles and similarly $\bft_m^-$ a set of $m$ holes excitations. The function $d(x,t; \bft_k^+, \bft_m^-)$ controls the space-time dependence of the correlation function and is given by
\begin{equation}
    d(x,t; \theta_1, \dots, \theta_k) = \sum_{j=1}^k \left(\omega(\theta_j^+)t - k(\theta_j^+)x\right) - \sum_{j=1}^m \left(\omega(\theta_j^-)t - k(\theta_j^-)x\right),
\end{equation}
with the integration measure defined by
\begin{equation}
    {\rm d} \bft_k^+ {\rm d} \bft_m^- = \frac{1}{k!} \frac{1}{m!}\prod_{j=1}^k \left(d\theta_j^+ \rho_{\rm h}(\theta_j^+)\right) \prod_{j=1}^m \left(d\theta_j^-  \rho_{\rm p}(\theta_j^-)\right).
\end{equation}
The functions $f_{\rho_{\rm p}}(\theta_1, \dots, \theta_k)$ are the thermodynamic form factors.  They are defined with reference to a function $\rho_{\rm p}(\theta)$ that describes a distribution function of particles in the finite energy density state.  It is uniquely related to the filling function $n(\theta)$ in a way that we describe later. The temporal-spatial dependence of the correlation functions depend upon $\omega(\theta_j)$ and $k(\theta_j)$, the energy and momentum carried by an excitation that reflect the presence of the particles in the finite energy density state.

In this work, we advance the thermodynamic bootstrap program. This program is a tool to obtain the form factors of local operators, $f_{\rho_{\rm p}}(\theta_1, \dots, \theta_m)$, in quantum integrable field theories, generalizing the vacuum bootstrap program~\cite{SmirnovBOOK,Karowski1978} to states at finite energy density.
The notion of thermodynamic form factors was first introduced in the context of matrix elements in the field theory corresponding to the quantum Ising model \cite{Leclair1996}.  This work was followed by a study \cite{Doyon2005} where it was argued that to account for thermal dressing effects, rapidity dependent leg factors were needed for form factors involving non-local operators.  Orthogonal to these approaches was the development of thermodynamic form factors for the non-relativistic Lieb-Liniger model \cite{Smooth_us,2018JSMTE..03.3102D,milosz_2021}. This work took the available determinantal expressions for the matrix elements and evaluated them in the thermodynamic limit.  More recently, simple thermodynamic form factors (ones involving one particle-hole excitation) were identified \cite{DoyonSpohn} as a result of generalized hydrodynamics \cite{doyon_ghd}.  Knowledge of two particle-hole form factor also play a central role in generalized hydrodynamics inasmuch as they determine behaviour beyond the Eulerian scale as first considered in~\cite{10.21468/SciPostPhys.6.4.049} and further reviewed in~\cite{GHD_review}.  Thermodynamic form factors recently have found application in formulating the non-equilibrium dynamics in weakly perturbed integrable models~\cite{weak_integ_breaking,PGK,denardis2022hydrodynamic,2023arXiv230312490L}.

Inspired by this initial work on TDFFs \cite{Leclair1996, Doyon2005,Smooth_us,2018JSMTE..03.3102D}, Ref. \cite{Bootstrap_JHEP} proposed
a set of axioms for integrable field theories with a single particle in their spectrum governing the TDFFs very much like the set of axioms that govern the vacuum form factors \cite{Karowski1978,SmirnovBOOK}.  The axioms proposed in \cite{Bootstrap_JHEP,Cortes_Cubero_2020} included a scattering axiom, a periodicity axiom, an annihilation pole axiom, and a normalization axiom.  

All of these axioms can be connected to the axioms governing the vacuum form factors in large, but finite volume.  As we elaborate on in Section 2, TDFFs can be thought of as a limit of a vacuum form factor with an appropriate number of particles.  The normalization axiom, which sets the value of the two-particle form factor, $f_{\rho_{\rm p}}(\theta_1, \theta_2)$ in the limit of $\theta_1$, approaching $\theta_2 + i\pi$, arises from a detailed calculation in Ref. \cite{Cortes_Cubero_2020} where an expression of this TDFF in terms of a finite volume vacuum form factor was developed.  The annihilation pole axiom provides allied information for TDFFs with a larger ($>2$) number of particle-hole excitations.  This axiom tells us that there will be simple poles in the TDFF as a function of one of the rapidities
whenever any particle approaches any hole and what the residue will be.  

While this axiom is natural, it was unclear whether the conjectured form of the axiom was consistent with its representation as a finite volume vacuum form factor.  In this work we will show that this is indeed the case. We achieve this in two steps. First, in Section~\ref{sec:residue} we use that the finite volume regularizations of TDFFs can be expressed through so-called almost diagonal vacuum form factors. For such vacuum form factors we prove a residue formula that mimics the annihilation pole axiom of the TDFFs. In the subsequent Section~\ref{sec:annihilation} we confirm that using this formula leads to the same residue as predicted by the annihilation pole axiom when limited to small excitations. This is the first main result of our work. As a direct consequence of this construction of the annihilation pole axiom, we are able to write down simple expressions for the TDFFs with multiple small particle-hole excitations.

The second result involves stating a crossing symmetry axiom for the TDFFs.
The TDFFs are formulated as matrix elements between a thermodynamic state with some number of finite excitations in the bra- (\emph{out}-) and ket- (\emph{in}-) state. In relativistic theory the particles in \emph{in-} and \emph{out}-states can be exchanged due to the crossing symmetry. The presence of the thermodynamic state breaks the relativistic invariance and therefore it is not clear whether some form of crossing symmetry survives. In this work we address this question and propose a crossing relation for the thermodynamic form factors. The form factors with crossed excitations are a necessary ingredient for ($n>2$)-pt correlation functions. We will consider this application in a separate publication. 

We argue for the crossing symmetry of the form factors in two ways. Firstly, in Section~\ref{sec:crossing_small}, we demonstrate it using the representation of the thermodynamic form factors through finite volume form factors. With the crossing symmetry at our disposal, in Section~\ref{sec:crossing_axiom} we formulate the TDFF axioms for generic form factors. Secondly, in Section~\ref{sec:gauge} we discuss the notion of \emph{reparameterization invariance}, the property that all thermodynamic quantities (not just TDFFs) should be invariant under the reparameterization of the thermodynamic state (up to subleading volume corrections).  We argue that from reparameterization invariance, the crossing relation follows naturally.

After establishing the form of crossing symmetry in the thermodynamic limit, we consider two applications. Firstly, in Section~\ref{sec:conserved_ff}, we write the crossed form factors of local charge and current operators associated with the conserved charges present in the field theory. Secondly, in~\ref{sec:non-relativistic} we use the relation between the relativistic sinh-Gordon theory and the non-relativistic Lieb-Liniger model~\cite{2009_Kormos_PRA_81,Kormos_2010} to compute the crossed form factors of the particle density operator in the latter. 

This work has four appendices. In Appendix~\ref{app:2pt} we sketch a derivation of the formula~\eqref{2point_fnc} for the two point from the finite volume expression.
In Appendix~\ref{app} we present an alternative derivation of the singular structure of the two particle-hole form factor based on finite volume regularization and explicit summation over singular contributions to it. In Appendix~\ref{app:extra_scattering}, we show further examples of the crossing symmetry for states consisting of few holes or few particle-hole excitations.
And finally in Appendix D, we briefly consider the possibilities of altering the minimal structure of the thermodynamic form factors through Castillejo–Dalitz–Dyson (CDD) that account for soft mode summation.

\section{Thermodynamic bootstrap program}

In this section we outline the thermodynamic bootstrap program.  We first provide details of the types of quantum field theories that we are studying here, both at zero temperature and finite energy density.  We then turn to the definition of the thermodynamic form factors and their representation in terms of finite volume vacuum form factors, a representation we use extensively in this paper.  Then finally, we turn to the set of axioms governing the TDFFs.

\subsection{Preliminaries: Setting the stage}

We consider an integrable quantum field theory (IQFT)~\cite{esskon,MussardoBOOK} with a single kind of particle which is simultaneously its own antiparticle. Representative examples of such theories include the sinh-Gordon field theory \cite{Vergeles:1976ra,MussardoBOOK} as well as certain massive perturbations of conformal minimal models \cite{Zamolodchikov1990}.  Examples also extend to theories with multiple particles provided one restricts oneself to states involving a single particle and the particle is its own anti-particle and does not have bound states with itself (i.e., certain sectors in the sine-Gordon field theory involving breathers).  One could also adapt easily the results herein to theories with differing particles and anti-particles provided one again restricts oneself to either the anti-particle- or particle-sector (i.e., sine-Gordon at one of its diagonal scattering points at zero temperature but finite chemical potential).   

The particles in the field theory have mass $m$ and their momentum and energy are parameterized in terms of the rapidity $\theta$,
\begin{equation} \label{bare_kinetics}
	e(\theta) = m \cosh \theta, \quad p(\theta) = m \sinh (\theta).
\end{equation}
The interaction processes in the IQFT are completely determined by the $2$-body scattering matrix $S(\theta)$ which is a function of the difference of rapidities. The S-matrix obeys the unitarity condition, 
\begin{equation}
    S(\theta)S(-\theta) = 1,
\end{equation}
and the crossing relation 
\begin{equation}
S(\theta) = S(i\pi - \theta).
\end{equation}
The S-matrix enters the commutation relations of the creation and annihilation operators of the particles, forming the Faddeev-Zamolodchikov algebra:
\begin{align}
    A(\theta_1) A(\theta_2) &= S(\theta_1 - \theta_2) A(\theta_2) A(\theta_1), \\
    A^{\dagger}(\theta_1) A^{\dagger}(\theta_2) &= S(\theta_1 - \theta_2) A^{\dagger}(\theta_2) A^{\dagger}(\theta_1), \\
    A(\theta_1) A^{\dagger}(\theta_2) &= S(\theta_2 - \theta_1) A^{\dagger}(\theta_2) A(\theta_1) + 2\pi \delta(\theta_2 - \theta_1).
\end{align}
The eigenstates of the system are then created by the successive action of the creation operator on the vacuum state $|0\rangle$,
\begin{equation}
    | \theta_1, \dots, \theta_n \rangle = A^{\dagger}(\theta_1) \cdots A^{\dagger}(\theta_n) |0\rangle,
\end{equation}
with $A(\theta) |0\rangle = 0$. The momentum and energy of the $n$-particle states are simply a sum of single-particle contributions, $p(\theta_j)$ and $e(\theta_j)$.

At finite energy density, the state of the system is characterized by a filling function $n(\theta)$. The interactive nature of the theory causes the density of particles, $\rho_{\rm p}(\theta)$, to depend on the presence of other particles. This leads to an integral equation for $\rho_{\rm tot}(\theta)$ involving the filling function~\cite{1969_Yang_JMP_10,Zamolodchikov1990,2012_Mossel_JPA_45},
\begin{equation}
	\rho_{\rm tot}(\theta) = \frac{1}{2\pi}p'(\theta) + \int {\rm d}\theta'\, T(\theta, \theta') n(\theta') \rho_{\rm tot}(\theta'),
\end{equation}
and the particle density is determined by $\rho_{\rm p}(\theta) = n(\theta) \rho_{\rm tot}(\theta)$. Here and at all points in the paper, integrals  $\int d\theta$ are to be understood over the whole real line.
The scattering kernel $T(\theta, \theta')$ is related to the $S$-matrix,
\begin{equation} \label{scattering_kernel}
	T(\theta, \theta') = \frac{1}{2\pi} \frac{\partial }{\partial \theta} \delta(\theta - \theta'), \qquad \delta(\theta)  = - i\log S(\theta).
\end{equation} 
The energy and momentum density of a state characterized by $\rho_{\rm p}(\theta)$ are given by
\begin{equation}
	\frac{E}{L} = \int {\rm d}\theta\, e(\theta) \rho_{\rm p}(\theta), \qquad \frac{P}{L} = \int {\rm d}\theta\, p(\theta) \rho_{\rm p}(\theta),
\end{equation}
where $L$ is the length of the $1d$ system which we assume to be large with respect to the inverse mass $m$. 
In general, the expectation value $Q$ of a local conserved charge $\hat{Q}$ takes form
\begin{equation}
	\frac{Q}{L} = \int {\rm d}\theta\, q(\theta) \rho_{\rm p}(\theta), 
\end{equation} 
where $q(\theta)$ is the eigenvalue of the charge on a single-particle state, $\hat{Q}|\theta\rangle = q(\theta)|\theta\rangle$. 

We define a finite density state $|\rho_{\rm p} \rangle$ as a state with an extensive number of particles distributed according to $\rho_{\rm p}(\theta)$. We can understand the state $|\rho_p\rangle$ in terms of a limiting process of finite volume states.  This will be described in detail in the next subsection.  With respect to this state we then consider particle and hole excited states. These are states where a finite number of particles are added or removed. We denote such states $|\rho_{\rm p}, \alpha_1, \dots, \alpha_m \rangle$ and call $\alpha_i$ excitations. In a relativistic theory the hole can be viewed as an antiparticle and antiparticles can be described by shifting its rapidity by $i\pi$. Therefore, if we allow the rapidities to be real or have imaginary parts equal to $i\pi$, this notation encompasses both particle and hole excitations.  

Because of the interactions present in the system, the excitations modify the distribution $\rho_{\rm p}(\theta)$ of the background particles. This modification is of order $1/L$ but affects every particle and therefore has a noticeable effect in thermodynamically large systems. The rapidity $\theta$ of a particle is shifted by an excitation $\alpha$ to a value $\theta'$ given by
\begin{equation} \label{shifted_theta}
    \theta' = \theta - \frac{F(\theta|\alpha)}{L \rho_{\rm tot}(\theta)},
\end{equation}
where the back-flow function $F(\theta|\alpha)$ obeys the following integral equation~\cite{1969_Yang_JMP_10},
\begin{equation} \label{backflow}
	F(\theta | \alpha) = \frac{1}{2\pi} \delta(\theta - \alpha) + \int {\rm d}\theta'\, T(\theta, \theta') n(\theta') F(\theta'|\alpha),
\end{equation}
and is additive in the excitations
\begin{equation}
	F(\theta | \{\alpha_j\}) = \sum_j F(\theta | \alpha_j).
\end{equation}
The back-flow function for holes obeys $F(\theta + i\pi| \alpha) = - F(\theta |\alpha)$ and $F(\theta|\alpha + i\pi) = - F(\theta|\alpha)$. These relations follow from the structure of the integral equation~\eqref{backflow} and analogous relations for $T(\theta, \alpha)$ which in turn follow from the unitarity and crossing symmetry of the $S$-matrix.\footnote{The unitarity and crossing symmetry of the $S$-matrix imply that the scattering phase obeys $\delta(\theta + i\pi) = - \delta(\theta)$.}

The expectation value of a conserved charge in an excited state has a contribution from the background particles and from the excitations. The latter takes into account the back-flow of the background particles,
\begin{equation}
	\langle \rho_{\rm p}, \alpha_1, \dots, \alpha_n | \hat{Q} | \rho_{\rm p}, \alpha_1, \dots, \alpha_n \rangle = \langle \rho_{\rm p}| \hat{Q} | \rho_{\rm p} \rangle + \sum_{j=1}^n q^{\rm Dr}(\alpha_j),
\end{equation}
where $q^{\rm Dr}(\alpha_j)$ is a renormalized single-particle eigenvalue. Following the convention from~\cite{PhysRevLett.124.140603}, we denote this renormalization by "Dressing" which is defined by \sout{the integral equation}
\begin{equation}\label{Dressing}
	f^{\rm Dr}(\theta) = f(\theta) - \int {\rm d}\theta'\, f'(\theta') n(\theta') F(\theta'| \theta). 
\end{equation}
The energy $\omega(k)$ and momentum $k(\theta)$ appearing in~\eqref{2point_fnc} are `Dressed' single-particle energy $e(\theta)$ and momenta $p(\theta)$ that appeared in eq.~\eqref{bare_kinetics}.

We also introduce the ``dressing" procedure (without the capitalization), as the solution to the following integral equation
\begin{equation}
	f^{\rm dr}(\theta) = f(\theta) + \int {\rm d}\theta' T(\theta, \theta') n(\theta') f^{\rm dr}(\theta').
\end{equation}
The two dressing procedures are related however through~\cite{PhysRevLett.124.140603},
\begin{equation}
	\frac{\partial f^{\rm Dr}(\theta)}{\partial \theta} = (f'(\theta))^{\rm dr}.
\end{equation}
We note that the back-flow function itself is proportional to the dressed scattering phase shift. This allows us to consider the effective scattering matrix\begin{equation}
	S^{\rm dr}(\theta, \theta') \equiv \exp\left(2 \pi i F(\theta| \theta') \right).
\end{equation}
Abusing terminology, we call $S^{\rm dr}(\theta, \theta')$ the dressed scattering matrix. However it is its logarithm of the S-matrix that is properly dressed.

Another important quantity for development of expressions for the TDFFs is the dressed differential scattering kernel $T^{\rm dr}(\theta, \theta')$ which is defined as the left dressing of $T(\theta, \theta')$. Explicitly, it obeys the following integral equation,
\begin{equation} \label{Tdr}
    T^{\rm dr}(\theta, \theta') = T(\theta, \theta') + \int_{-\infty}^{\infty}{\rm d}\mu\, T(\theta, \mu) n(\mu) T^{\rm dr}(\mu, \theta').
\end{equation}
The dressed differential scattering kernel is related to the back-flow,
\begin{equation}
    T^{\rm dr}(\theta, \alpha) = - \frac{\partial F(\theta|\alpha)}{\partial \alpha}.
\end{equation}
The background distribution, $\rho_{\rm p}(\theta)$, the back-flow function $F(\theta | \alpha)$, and the dressing procedures described here characterize the thermodynamics of finite-energy density states in the IQFT and excitations around it.  With these preliminaries complete, we turn now to the definition of the TDFFs.

\subsection{Definition of thermodynamic form factors}

The TDFFs, $f_{\rho_p}$, are defined as matrix elements of a local operator, $\mathcal{O}(0,0)$, between the two thermodynamic states.  We introduce the following notation for TDFFs with respect to a finite energy density state described by the particle distribution, $\rho_p(\theta)$ 
\begin{equation} \label{def_ffn}
	f_{\rho_{\rm p}}(\alpha_1, \dots, \alpha_m) = \frac{L^{m/2}\langle \rho_{\rm p} | \mathcal{O}(0,0) | \rho_{\rm p}, \alpha_1 \dots, \alpha_m \rangle}{\sqrt{\langle \rho_{\rm p} | \rho_{\rm p}\rangle \langle \rho_{\rm p}, \alpha_1 \dots, \alpha_m  | \rho_{\rm p}, \alpha_1 \dots, \alpha_m \rangle}}.
\end{equation}
The general case of a TDFF involving an operator $\mathcal{O}(x,t)$ at a non-zero spacetime point is then obtained via 
\begin{equation}
	\frac{L^{m/2}\langle \rho_{\rm p} | \mathcal{O}(x,t) | \rho_{\rm p}, \alpha_1 \dots, \alpha_m \rangle}{\sqrt{\langle \rho_{\rm p} | \rho_{\rm p}\rangle \langle \rho_{\rm p}, \alpha_1 \dots, \alpha_m  | \rho_{\rm p}, \alpha_1 \dots, \alpha_m \rangle}} = e^{i x \sum_{j=1}^m k(\alpha_j) - i t \sum_{j=1}^m \omega(\alpha_j)}  f_{\rho_{\rm p}}(\alpha_1, \dots, \alpha_m),
\end{equation}
where $k(\alpha)$ and $\omega(\alpha)$ are Dressed (as defined in eq.~\ref{Dressing}) bare momenta $p(\alpha)$ and bare energies $e(\alpha)$ of~\eqref{bare_kinetics} respectively.  

To anchor this notation, we now define the TDFFs in terms of finite volume vacuum form factors of an extensive number of particles.  This entails three steps.  We first need to define what we mean by a thermodynamics state $|\rho_p\rangle$ in terms of finite volume states.  Secondly, we need to define the normalization of these states as it is an ingredient to eq. \ref{def_ffn}.  Finally we need to state how to identify the matrix element $\langle \rho_{\rm p}|\mathcal{O}(0,0)| \rho_{\rm p}, \alpha_1 \dots, \alpha_m \rangle$ with a corresponding matrix element in the finite volume theory.

To take the first step, we begin by defining n-particle states in finite volume.  Such a state in finite volume can be written as
\begin{equation}
\vert I_1,\dots,I_n\rangle =  \frac{1}{\sqrt{\rho_n(\theta_1,\cdots,\theta_n)}}A^\dagger (\theta_1) \cdots A^\dagger (\theta_n)|0\rangle \equiv \frac{1}{\sqrt{\rho_n(\theta_1,\cdots,\theta_n)}}|\theta_1,\cdots,\theta_n\rangle,
\end{equation}
where the $A^\dagger$'s are the finite volume version of the Faddeev-Zamolodchikov operators.
The numbers $\{I\}$ here are a set of integers (or half-integers, depending on the particle statistics).  The integers determine the rapidities of the n-particles via the Bethe equations~\cite{MussardoBOOK}
\begin{equation} \label{Bethe}
Q_k(\theta_1,\dots,\theta_n)\equiv L m\sinh\theta_k+\sum_{l\neq k}\delta(\theta_k-\theta_l)=2\pi I_k, \qquad k = 1, \dots, n,
\end{equation}
which in turn depend on the two-body scattering phase $\delta(\theta)$ of the theory.  The Bethe equations then serve as a map from a set of quantum numbers $\{I\}$ to a set of of rapidities $\{\theta\}$. 

The normalization of a finite volume state is given by
\begin{eqnarray} \label{norm_finite} 
	\langle \theta_1,\cdots,\theta_n |\theta_1,\cdots,\theta_n \rangle &=& \rho_{n}(\theta_1, \dots, \theta_n) \equiv \vert {\rm Det} J(\theta_1,\dots,\theta_n)\vert,
 \end{eqnarray}
 where
 $J$ is a matrix whose elements are defined by
 \begin{eqnarray}
 J_{kl}(\theta_1,\dots,\theta_n) = \frac{\partial}{\partial\theta_l}Q_k(\theta_1,\dots,\theta_n).
\end{eqnarray}
This determinant can be rewritten as a determinant of the scattering kernel $T$.

With this definition of the finite volume states in hand, we can now define what we mean by thermodynamic states precisely.  We understand the thermodynamic state $|\rho_p\rangle$ as a certain limit of finite volume states.  Consider a finite volume state at volume $L$ constructed by choosing a set of integers $\{I_i\}_L$ which, through the Bethe equations \ref{Bethe}, determine a set of $\{\theta_i\}_L$'s whose distribution provides a discrete approximation of $\rho_p(\theta)$. To see what we mean by ``provides a discrete approximation", divide the real axis into a set of intervals $(\beta_j,\beta_{j+1})$, $j=-\infty,\cdots,\infty$ of width $\Delta\beta$.  Then write the number of rapidities in the set $\{\theta_i\}$ in interval $(\beta_j,\beta_{j+1})$ as $N_j$.
We want to choose the set $\{I_i\}$ so that
\begin{equation}
    N_j \approx L \rho_p(\beta_j)\Delta\beta .
\end{equation}
We then want to consider of sequence of such sets $\{\theta_i\}_L$ at ever increasing $L$ such that we can take a limit where
\begin{equation}
\rho_p (\beta_j) = \lim_{L\rightarrow\infty,\Delta\beta\rightarrow 0} N_j/(\Delta \beta L).
\end{equation}
We will denote this ``thermodynamic" limit henceforth as $\lim_{td}$:
\begin{equation}
    \lim_{td} \equiv \lim_{n\rightarrow\infty,L\rightarrow\infty,\Delta\beta\rightarrow 0}.
\end{equation}
The thermodynamic state is then identified with this limiting sequence
\begin{equation}
    |\rho_p\rangle = \lim_{td} \sqrt{\rho_n(\theta_1, \cdots, \theta_n)}|I_1,\cdots,I_n\rangle.
\end{equation}
It correspondingly has a normalization 
\begin{equation}
    \langle \rho_p|\rho_p \rangle = \lim_{td} \rho_n(\theta_1,\cdots,\theta_n) = \left(\lim_{td} \prod_{j=1}^n \left( 2 \pi L \rho_{\rm tot}(\theta_j)\right)\right) \times \det\left(1 - Tn\right) (1 + \mathcal{O}(1/L)),
\end{equation}
where the determinant is the Fredholm determinant of the kernel $T(\theta, \theta') n(\theta')$~\cite{KorepinBOOK}. The details of this are not important for us. The only relevant aspect is that the determinant is finite in the thermodynamic limit and depends on the state only through function $n(\theta)$. The norm itself is diverging in the thermodynamic limit, but when combined with other factors entering the definition of the thermodynamic form-factors, the expression will be finite.  

We are now in a position to give the normalization of the thermodynamic excited state.  Supposing we have $k$-particle excitations, $\{\alpha_i\}$, we can define the thermodynamic excited state
\begin{align}
|\rho_p,\alpha_1,\cdots,\alpha_k\rangle &\equiv \lim_{td} A^\dagger(\theta_1')\cdots A^\dagger(\theta_{n}') A^\dagger(\alpha_1)\cdots A^\dagger({\alpha_k})|0\rangle \nonumber \\
& \equiv \lim_{td} \rho_{n+k}(\theta'_1,\cdots,\theta'_n,\alpha_1,\cdots,\alpha_k)^{1/2}  \times|I_1,\cdots,I_n,I_{\alpha_1},\cdots,I_{\alpha_k}\rangle,
\end{align}  
with $I_{\alpha_j}$ the quantum number corresponding to the rapidity $\alpha_j$.
Here the $\lim_{td}$ involves taking the number of particles describing $\rho_p$ to $\infty$ with $k$ fixed.  We have placed primes on the rapidities $\theta'$'s in this excited state as the excitations alters the quantization condition of the $\theta$'s in comparison to the base thermodynamic state, see~\eqref{shifted_theta}. In writing the excited thermodynamic state we have made certain choices of which the reader should be aware.  The creation operators creating the excitations in the finite volume state come before the creation operators creating the base thermodynamic state $|\rho_p\rangle$.

The normalization of the excited thermodynamic state is then given by
\begin{eqnarray}
    &&\langle \alpha_k,\cdots,\alpha_1,\rho_p|\rho_p,\alpha_1,\cdots,\alpha_k\rangle = \lim_{td} \rho_{n+k}(\theta'_1,\cdots,\theta'_n,\alpha_1,\cdots,\alpha_k)\cr\cr
    && = \langle \rho_p|\rho_p \rangle  \lim_{td} \left(\prod^n_{i=1}\left(1+(\theta_i'-\theta_i)\frac{\rho'_{tot}(\theta_i)}{\rho_{tot}(\theta_i)}\right)\prod^k_{i=1} \left(2\pi L \rho_{tot}(\alpha_i)\right) \right)
\end{eqnarray}
where
\begin{eqnarray}\label{kdef}
 \theta_i'-\theta_i &=&  - \frac{1}{L\rho_{tot}(\theta_i)}\sum^k_{j=1}F(\theta_i|\alpha_j);\cr\cr
   \prod^n_{i=1}\left( 1+(\theta_i'-\theta_i)\frac{\rho'_{tot}(\theta_i)}{\rho_{tot}(\theta_i)}\right)&=&\prod^k_{j=1} K(\alpha_j) + \mathcal{O}(\frac{1}{L}) \cr\cr 
   K(\alpha) &\equiv& \exp\bigg[-\int d\theta n(\theta)\frac{\rho'_{tot}(\theta)}{\rho_{tot}(\theta)} F(\theta|\alpha)\bigg] .
\end{eqnarray}
In this expression for the normalization we have taken into account the effects of the shifted $\theta'$'s due to the presence of excitations.  This effect can be written in a compact fashion as a product indexed by the different excitations:
\begin{equation}\label{norm}
     \langle \alpha_k,\cdots,\alpha_1, \rho_p|\rho_p,\alpha_1,\cdots,\alpha_k\rangle = \langle \rho_p|\rho_p \rangle \prod^k_{i=1} N(\alpha_i), ~~~ N(\alpha) \equiv 2\pi L|\rho_{tot}(\alpha)|K(\alpha).
\end{equation} 

Having defined the norms of the thermodynamic states, we are now in a position to move on to defining the thermodynamic form factors in terms of the form factors in finite volume, $f_{FV}$.  The finite volume form factors are defined as the matrix elements
\begin{equation}
    f_{FV}(I_1',\dots,I_m'|I_1,\dots,I_n\rangle \equiv \langle I_1^\prime,\dots,I_m^\prime\vert O(0)\vert I_1,\dots,I_n\rangle .
\end{equation} 
The finite volume form factors can be expressed in terms of vacuum form factors, $f_{IV}$, in {\it infinite} volume.  Writing the infinite volume form factors as 
\begin{equation} \label{vacuum_ff}
    f_{IV}(\theta_1, \dots, \theta_n) = \langle 0| \mathcal{O}(0) | \theta_1, \dots, \theta_n \rangle = \langle 0| \mathcal{O}(0) A^\dagger(\alpha_1) \dots, A^\dagger(\alpha_n) |0\rangle,
\end{equation}
we can follow the work of Refs.~\cite{Pozsgay2008a,Pozsgay2008b} to write $f_{FV}$ in terms of $f_{IV}$ via :
\begin{equation} \label{finite_volume_ff}
f_{FV}(I_1',\dots,I_m'|I_1,\dots,I_n) = \frac{f_{IV}(\theta_1^\prime+\pi{\rm  i},\dots,\theta_m^\prime+\pi {\rm i},\theta_1,\dots,\theta_n)}{\sqrt{\rho_m(\theta_1^\prime,\dots,\theta_m^\prime)\rho_n(\theta_1,\dots,\theta_n)}}+\mathcal{O}(e^{-mL}),
\end{equation}
This expression in eq.~\ref{finite_volume_ff} is accurate up to corrections $\mathcal{O}(e^{-mL})$ that disappear rapidly in large volumes. The advantage of this expression is that the finite volume form factors can be understood in terms of their infinite volume counterparts for which we have good control because of the (vacuum) form factor bootstrap.   

With FVFFs defined we are finally in a position to write the TDFFs in terms of FVFFs (and their representation in terms of infinite volume FFs). Suppose we have a set of $k$ particle excitations at $\alpha_1, \cdots, \alpha_k$ and a set of $l$ hole excitations at $\gamma_1,\cdots,\gamma_l$.  In order to define the TDFF, we need to consider the sequence of states with n particles $|\theta_1,\cdots,\theta_n\rangle$ whose limit $n \rightarrow \infty$ defines the reference state $\rho_p(\theta)$.  For hole excitations, we suppose that each such state possesses the rapidities at which we want to make the holes, i.e., $\theta_{n+1-i}=\gamma_{i},~~i=1,\cdots,l$.  The TDFF is then defined by 
\begin{eqnarray} \label{general_fv}
	f_{\rho_{\rm p}}(\gamma_1+i\pi,\cdots,\gamma_l+i\pi,\alpha_1, \dots, \alpha_k) && \cr\cr
&& \hskip -2.5in = \lim_{td} L^{(k+l)/2} \langle I_{\gamma_1},\cdots,I_{\gamma_l}, I_{n-l},\cdots,I_1|\mathcal{O}|I_1,\cdots,I_{n-l} ,I_{\alpha_1},\cdots,I_{\alpha_k}\rangle
\cr\cr 
&&\hskip -2.5in = \lim_{td} L^{(k+l)/2} \frac{f_{IV}(\gamma_1+i\pi,\cdots,\gamma_l+i\pi, \theta_{n-l}+\pi{\rm  i},\dots,\theta_1+\pi {\rm i},\theta'_1,\cdots,\theta'_{n-l},\alpha_1,\cdots,\alpha_k)}{\big(\rho_{n}(\theta_1,\dots,\theta_{n-l}, \gamma_l,\cdots,\gamma_1)\rho_{n+k-l}(\theta'_1,\cdots,\theta'_{n-l},\alpha_1,\cdots,\alpha_k)\big)^{1/2}} \cr\cr
&&\hskip -2.5in =  \lim_{td} L^{(k+l)/2} \frac{f_{IV}(\gamma_1+i\pi,\cdots,\gamma_l+i\pi, \theta_{n-l}+\pi{\rm  i},\dots,\theta_1+\pi {\rm i},\theta'_1,\cdots,\theta'_{n-l},\alpha_1,\cdots,\alpha_k)}{\rho_{n}(\theta_1,\dots,\theta_{n-l}, \gamma_l, \dots, \gamma_1) \sqrt{\prod N(\alpha_i) \prod N^{-1}(\gamma_i )}}\cr &&,
\end{eqnarray}
where the background rapidities $\{\theta_j'\}$ of the excited state are shifted due to the presence of $\{\alpha_i\}$ particles and the absence of $\{\gamma_i\}$ particles,
\begin{equation}
    \theta_j' = \theta_j + \sum^l_{j=1}F(\theta_i|\gamma_j) - \sum^k_{j=1}F(\theta_i|\alpha_j).
\end{equation}
In the definition of the thermodynamic form factor we include a power of the system size, $L^{(k+l)/2}$.  This factor arises because we are absorbing a phase space factor into the definition of the form factor that will allow us to write correlation functions in terms of TDFF as in eqn.~\ref{2point_fnc} (see Appendix~\ref{app:2pt} for the details).
In writing~\eqref{general_fv} we make sure to order the hole excitations so that $f_{\rho_{\rm p}}(\gamma_1+i\pi,\cdots,\gamma_l+i\pi,\alpha_1, \dots, \alpha_k)$ will obey the scattering axiom, i.e.,
\begin{eqnarray}
    && f_{\rho_{\rm p}}(\gamma_1+i\pi,\gamma_2+i\pi,\cdots,\gamma_l+i\pi,\alpha_1, \dots, \alpha_k) \cr\cr 
    &&\hskip 1in = S(\gamma_1-\gamma_2)f_{\rho_{\rm p}}(\gamma_2+i\pi,\gamma_1+i\pi,\cdots,\gamma_l+i\pi,\alpha_1, \dots, \alpha_k).
\end{eqnarray}
We note also that in comparison to the previous work on TDFFs~\cite{Bootstrap_JHEP,Cortes_Cubero_2020}, we have made a change in the normalization of the form factors in this work. Previously, they were normalized with respect to the background state only, whereas now they are normalized with respect to both the base state {\it and} its excited counterpart. We have made this change for two reasons. Firstly, this normalization makes it more convenient to formulate crossing symmetry. Secondly, this definition aligns with the one used for the thermodynamic form factors in the Lieb-Liniger model~\cite{Smooth_us}.

We call the expression~\eqref{general_fv} for the TDFF its finite volume regularization. It can be written in a more symmetric way by extracting the holes contribution from the normalization of the background state and looking at a reference state described in finite volume with $n+l$ rapidities. We have 
\begin{eqnarray} \label{general_fv2}
	f_{\rho_{\rm p}}(\gamma_1+i\pi,\cdots,\gamma_l+i\pi,\alpha_1, \dots, \alpha_k) && \cr\cr
&& \hskip -2.5in =  \lim_{td} L^{(l+k)/2} \frac{f_{IV}(\gamma_1+i\pi,\cdots,\gamma_l+i\pi, \theta_n+\pi{\rm  i},\dots,\theta_1+\pi {\rm i},\theta'_1,\cdots,\theta'_n,\alpha_1,\cdots,\alpha_k)}{\rho_{n}(\theta_1,\dots,\theta_n) \sqrt{\prod N(\alpha_j) \prod N(\gamma_j)}},\quad
\end{eqnarray}
where we used that
\begin{equation}
\rho_{n+l}(\theta_1,\dots,\theta_n, \gamma_1, \dots, \gamma_l) = \rho_n(\theta_1, \dots, \theta_n) \prod_{j=1}^l N(\gamma_j) \times \left( 1 + \mathcal{O}(1/L)\right),
\end{equation}
which follows from~\eqref{norm_finite} and~\eqref{norm}.

\subsection{Thermodynamic form factor axioms}

Having defined both our theory at finite energy density (Section 2.1) and our definition of the thermodynamic form factors (Section 2.2), we are now in position to state the axioms for the TDFFs.
They are as follows \cite{Bootstrap_JHEP,Cortes_Cubero_2020}:
\begin{itemize}
	\item Scattering axiom
	\begin{equation} \label{scattering_axiom}
		f_{\rho_{\rm p}}(\alpha_1, \dots, \alpha_j, \alpha_{j+1}, \dots, \alpha_m) = S(\alpha_j - \alpha_{j+1}) f_{\rho_{\rm p}}(\alpha_1, \dots, \alpha_{j+1}, \alpha_{j}, \dots, \alpha_m);
	\end{equation}
	\item Periodicity axiom
	\begin{equation} \label{periodicity_axiom}
		f_{\rho_{\rm p}}(\alpha_1, \dots, \alpha_m) = R_{\rho_{\rm p}}(\alpha_m| \alpha_1, \dots, \alpha_m) f_{\rho_{\rm p}}(\alpha_m + 2 i\pi , \alpha_1, \dots, \alpha_{m-1});
	\end{equation}
	\item Annihilation pole axiom %
	\begin{align}
		- i \res{\alpha_1 = \alpha_2 }f_{\rho_{\rm p}}&(\alpha_1 + i\pi, \alpha_2, \dots, \alpha_m) \nonumber \\
  & = \frac{1 - R_{\rho_{\rm p}}(\alpha_2| \alpha_3, \dots, \alpha_m) \prod_{j=3}^m S(\alpha_2 - \alpha_j)}{2\pi \rho_{\rm tot}(\alpha_1)} f_{\rho_{\rm p}}(\alpha_3, \dots, \alpha_m); \label{annihilation_axiom}
	\end{align}
	\item Normalization axiom
	\begin{equation} \label{normalization_axiom}
		\lim_{\kappa \rightarrow 0}f_{\rho_{\rm p}} (\alpha + i \pi, \alpha + \kappa) = V_{\rho_{\rm p}}(\alpha).
	\end{equation}
\end{itemize}
In comparison with the axioms as stated in Refs.~\cite{Bootstrap_JHEP,Cortes_Cubero_2020}, the new normalization, i.e. eq.~\ref{def_ffn}, only affects the annihilation pole axiom and the normalization axiom. To obtain the axioms as originally stated, the right-hand sides of the equations of the two axioms must be multiplied by $2\pi \rho_{\rm tot}(\alpha_1)$ and $2\pi \rho_{\rm tot}(\alpha)$, respectively.
Although the list of axioms also includes a cluster property and a bound on growth (see~\cite{Bootstrap_JHEP}), we will not be using them in this work and hence do not include them here. The axioms involve the (bare) scattering matrix $S(\theta)$ and the ratio between the dressed scattering matrix and the bare one, 
\begin{equation} \label{def_R}
	R_{\rm p}(\alpha|\alpha') = \frac{S^{\rm dr}(\alpha,\alpha')}{S(\alpha - \alpha')} = \exp\left(2\pi i F(\alpha|\alpha') - i \delta(\alpha - \alpha') \right).
\end{equation}
For multiple excitations we have
\begin{equation}
	R_{\rm p}(\alpha|\alpha_1, \dots, \alpha_m) = \prod_{j=1}^m R_{\rho_{\rm p}}(\alpha | \alpha_j).
\end{equation}
The normalization axiom involves the function $V_{\rho_{\rm p}}(\alpha)$ defined through a LeClair-Mussardo like series~\cite{2018JSMTE..03.3102D,Cortes_Cubero_2020}, 
\begin{eqnarray}\label{V_function}
	V_{\rho_{\rm p}}(\alpha) &=& \lim_{\kappa\rightarrow 0}f_{\rho_{\rm p}}(\alpha + i \pi, \alpha + \kappa) \cr\cr 
 &=&\frac{1}{2\pi \rho_{\rm tot}(\alpha)} \sum_{k=0}^{\infty} \frac{1}{k!}\int \prod_{j=1}^k \left( \frac{d\theta_j}{2\pi} n(\theta_j) \right) f_{IV,c}(\theta_1, \dots, \theta_k, \alpha).
\end{eqnarray}
In this expression $f_{IV,c}(\theta_1, \dots, \theta_m)$ is the connected infinite volume vacuum form factor which we define in Section~\ref{sec:residue}.
Finally, for the residue, appearing in the annihilation pole axiom, we use the following convention
\begin{equation}
	\res{x=a} f(x) = \lim_{x\rightarrow 0} (x-a) f(x).
\end{equation}
In the limit of vanishing energy density, $\rho_{\rm p} \rightarrow 0$, the axioms reduce to those of the vacuum form factors~\cite{SmirnovBOOK}. In this limit, we have $R_{\rm p}(\alpha|\alpha') \rightarrow 1$, and the periodicity axiom and annihilation pole axioms above reduce to their standard form. 

Via the annihilation pole axiom, the TDFFs with more particle-hole excitations contain simple poles.  In the next two sections we verify the annihilation pole axiom is consistently stated in terms of the representation of the TDFFs in terms of their finite volume counterparts.  We show this in two steps. Firstly, in Section 3 we derive a general residue formula for the so-called almost diagonal form factors.  Such form factors are infinite volume form factors where the rapidities in the \emph{in}- and \emph{out}-states are slightly displaced from one another.  This residue formula is valid independent of any discussion of TDFFs and stands as an independent result of our work. Secondly, in Section~\ref{sec:annihilation}, we use this formula in the context of the finite volume form factor representation of the TDFFs and derive the annihilation pole axiom in the limit of small excitations.

\section{Residue formula for almost diagonal vacuum form factors} \label{sec:residue}

In this section we take the first step in validating the annihilation pole axiom for the TDFFs.  In this first step we consider infinite volume vacuum form factors and find the residue formula for a particular limit of the infinite volume form factors, what we call almost diagonal form factors (defined directly below). While this result is important for deriving the annihilation pole axiom for the thermodynamic form factors, it also is a result that stands on its own. It is an exact result about the analytic structure of vacuum form factors. 

The almost diagonal form factors are defined as
\begin{equation} \label{almost_diagonal_definition}
    f_{IV,d}(\theta_1, \dots, \theta_n| \epsilon_1, \dots, \epsilon_n) = f_{IV}(\theta_n + \pi i , \dots, \theta_1 + \pi i, \theta_1 + \epsilon_1, \dots, \theta_n + \epsilon_n), 
\end{equation}
where the $\epsilon_i$'s are to be thought of as small, that is, the form factor is sitting near its (vacuum) annihilation poles.
If we extract the terms singular in $\epsilon_i$, we find that they have the following structure~\cite{Pozsgay2008b},
\begin{equation}
	f_{IV,d}(\theta_1, \dots, \theta_n| \epsilon_1, \dots, \epsilon_n) = \frac{1}{\epsilon_1 \cdots \epsilon_n} \sum_{i_1 = 1}^n \cdots \sum_{i_n = 1}^n a_{i_1, \dots, i_n} (\theta_1, \dots, \theta_n)\, \epsilon_{i_1} \cdots \epsilon_{i_n} + (\dots). \label{almost_diagonal_singular}
\end{equation}
Here $a_{i_1, \dots, i_n}$ are finite and symmetric in the indices and $(\dots)$ denotes terms that are finite in the simultaneous limit $\epsilon_i \rightarrow 0$ of all the $\epsilon$'s. 
The connected (infinite volume) form factor is now defined as the $\epsilon_i$-independent part of the almost diagonal form factor, namely
\begin{equation}
	f_{IV,c}(\theta_1, \dots, \theta_n) = n!\, a_{1,\dots, n}(\theta_1, \dots, \theta_n).
\end{equation}
$f_{IV,c}$ is a symmetric function of its variables. 

The singular structure appearing in~\eqref{almost_diagonal_singular} can be efficiently described as a sum over certain graphs as proven in~\cite{Pozsgay2008b}. We recall now this result. Schematically, the formula is
\begin{equation} \label{Balazs_graphs}
	f_{IV,d}(\theta_1, \dots, \theta_n| \epsilon_1, \dots, \epsilon_n) = \sum_{g \in G_n} h(g),
\end{equation}
where $G_n$ is a set of certain graphs and $h(g)$ is a function which associates to a graph $g$ its contribution to the almost diagonal form factor. We discuss now the details of this equation.

\begin{figure}
    \centering
    \includegraphics[scale=0.6]{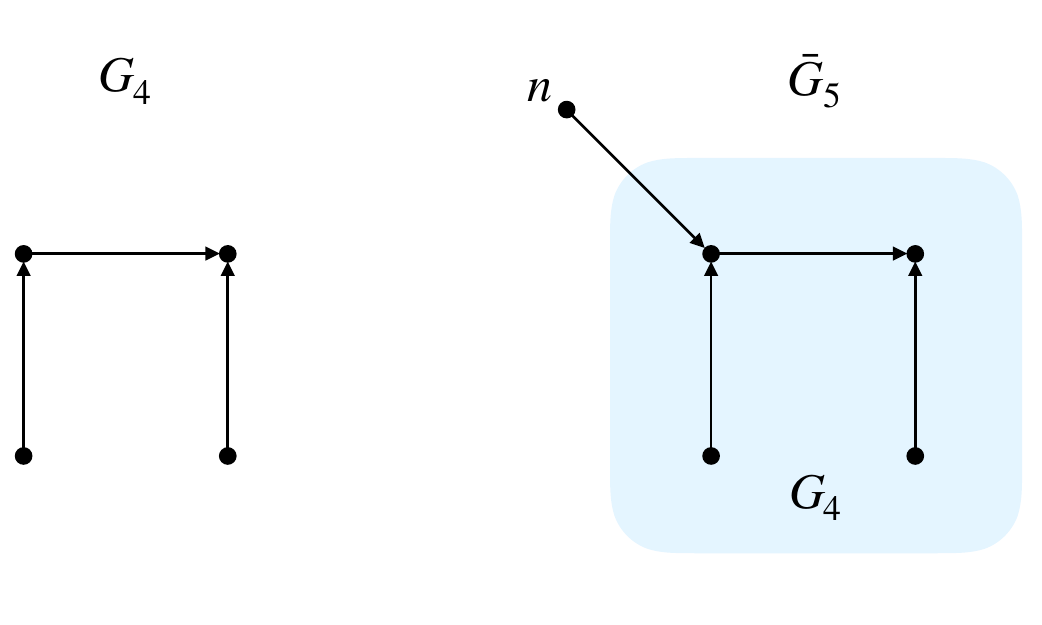}
    \caption{(left panel) An example of a graph belonging to $G_4$. The graph is tree-like (there are no loops) and there is at most one outgoing edge from any vertex. (right panel) An example of a graph belonging to $\bar{G}_5$. It contains a single edge from vertex $n=5$ to one of the vertices of a graph belonging to $G_4$.}
    \label{fig:graphs}
\end{figure}

$G_n$ is a set of directed graphs with $n$ vertices such that each graph $g \in G_n$ is tree-like and for each vertex there is at most one outgoing edge, see Fig.~\ref{fig:graphs}. We denote as $E_{jk}$ an edge in a graph going from node $j$ to node $k$.  With each node of the graph $g$, we associate a rapidity and this assignment splits the set of rapidities in two sets $\{\theta\} = \{\theta^+\} \cup \{\theta^-\}$ with $\{\theta^+\}$ associated to the nodes without an outgoing arrow and $\{\theta^-\}$ to the nodes with outgoing arrows. Then a graph $g$ contributes a factor $f_{IV,c}(\{\theta^+\})$ multiplied by for each edge, $E_{jk}$, $2\pi (\epsilon_k/\epsilon_j) T(\theta_j - \theta_k)$. This product defines the function $h(g)$ and completes the description of eq.~\eqref{Balazs_graphs}.

The aim of this section is to analyze the divergence when one of the $\epsilon_i$'s approaches $0$. Namely, we want to compute
\begin{equation}
	\res{\epsilon_n = 0}\, f_{IV,d}(\theta_1, \dots, \theta_n| \epsilon_1, \dots, \epsilon_n).
\end{equation}
We start by analysing the dependence of $f_{IV}(\theta_n + \pi i , \dots, \theta_1 + \pi i, \theta_1 + \epsilon_1, \dots, \theta_n + \epsilon_n)$ on $\epsilon_n$. Contributions to this expression come from a sum over different graphs. According to the rules stated above, a contribution from graph $g \in G_n$ is proportional to $\epsilon_n^{N_i - N_o}$ where $N_i$ is the number of edges ending at the node $n$ (incoming edges) and $N_o$ is the number of outgoing edges from the same node $n$. Because possible values for $N_o$ are $0$ and $1$, only graphs for which $N_i=0$ and $N_o=1$ contribute to the residue. We denote such graphs by $\bar{G}_n$.

Each graph $g$ in $\bar{G}_n$ contains a single edge from the node $n$ to some other node, see Fig.~\ref{fig:graphs}. There is no other edge connected to node $n$. Therefore each such graph can be uniquely obtained by adding an edge $E_{nk}$ to some graph $g' \in G_{n-1}$ with $k=1, \dots, n-1$ referring to each of its nodes. Note that adding an arrow from the $n$-th node to the $k$-th node is always possible and does not change the character of the $k$-th node. Therefore
\begin{equation}
	h(E_{nk} + g') = 2\pi \frac{\epsilon_k}{\epsilon_n} T(\theta_n - \theta_k) \times h(g'),
\end{equation}
and for the almost diagonal form factor we can isolate the terms contributing as $\epsilon^{-1}_n$ as follows
\begin{equation}
	f_{IV,d}(\theta_1, \dots, \theta_n| \epsilon_1, \dots, \epsilon_n) = 2\pi \sum_{k=1}^{n-1} \frac{\epsilon_k}{\epsilon_n} T(\theta_n - \theta_k) \times \sum_{g' \in G_{n-1}}h(g') + (\dots),
\end{equation}
where $(\dots)$ denote terms regular in $\epsilon_n$. The remaining sum over the graphs gives the almost-diagonal form factor of $2(n-1)$ particles and therefore
\begin{equation}
	f_{IV,d}(\theta_1, \dots, \theta_n| \epsilon_1, \dots, \epsilon_n) = 2\pi \left( \sum_{j=1}^{n-1} \frac{\epsilon_j}{\epsilon_n} T(\theta_n - \theta_j) \right)\times f_{IV,d}(\theta_1, \dots, \theta_{n-1}| \epsilon_1, \dots, \epsilon_{n-1}) + (\dots).
\end{equation}
We can also state this result in the residue notation as
\begin{equation}
	\res{\epsilon_n = 0}\, f_{IV,d}(\theta_1, \dots, \theta_n| \epsilon_1, \dots, \epsilon_n) = 2\pi \left( \sum_{j=1}^{n-1} \epsilon_j T(\theta_n - \theta_j) \right) f_{IV,d}(\theta_1, \dots, \theta_{n-1}| \epsilon_1, \dots, \epsilon_{n-1}).
\end{equation}
Nothing in our derivation depends on the ordering of the rapidities, therefore, we can generalize it to
\begin{eqnarray}\label{fIVd}
	\res{\epsilon_i = 0}\, f_{IV,d}(\theta_1, \dots, \theta_n| \epsilon_1, \dots, \epsilon_n) && \cr\cr 
 && \hskip -1.5in =2\pi \left( \sum_{\substack{j=1 \\ j \neq i}}^n \epsilon_j T(\theta_n - \theta_j) \right) f_{IV,d}(\theta_1, \dots, \hat{\theta}_i, \dots, \theta_n| \epsilon_1, \dots, \hat{\epsilon}_i, \dots,\epsilon_n),
\end{eqnarray}
where with the hat symbol we denote the absence of a given variable. This is the final result of this section.

\section{The annihilation pole axiom for thermodynamic form factors in the limit of low energy particle-hole excitation} \label{sec:annihilation}

We will now show the consistency between the finite-volume regularization of the thermodynamic form factors and the TDFF annihilation pole axiom in the limit of small excitations. The former uses the notion of almost diagonal form factors and we will use the residue formula derived in the previous section. 
One consequence of our validation of the annihilation pole axiom will be that we will have obtained valid formulae for the TDFFs involving multiple particle-hole excitations in the low energy limit.  This extends what we know about such TDFFs from the one and two particle-hole cases.

We start with rewriting the annihilation pole axiom for low energy particle-hole excitations.  We recall the annihilation pole axiom from~\eqref{annihilation_axiom} is given by,
\begin{align}
	- i \res{\alpha_1 = \alpha_2} f_{\rho_{\rm p}} &(\alpha_1 + i \pi, \alpha_2, \alpha_3, \dots, \alpha_n) = \nonumber \\
 & = \frac{1 - R_n(\alpha_2| \alpha_3, \dots, \alpha_n) \prod_{j=3}^n S(\alpha_2 - \alpha_n)}{2\pi \rho_{\rm tot}(\alpha_1)}f_{\rho_{\rm p}}(\alpha_3, \dots, \alpha_n).
\end{align}
We note that function $R_n(\alpha|\alpha')$ defined in eq.~\eqref{def_R} can be written as follows
\begin{equation}
	R_n(\alpha | \alpha') = e^{2 \pi i F(\alpha| \alpha')} S^{-1}(\alpha - \alpha').
\end{equation}
This allows us to express the annihilation pole axiom solely in terms of the back-flow function
\begin{align}
	- i \res{\alpha_1 = \alpha_2} f_{\rho_{\rm p}} (\alpha_1 + i \pi, \alpha_2, \alpha_3, \dots, \alpha_n) = \frac{1 - \exp\left( 2\pi i \sum_{j=3}^n F(\alpha_2| \alpha_j) \right)}{2\pi \rho_{\rm tot}(\alpha_1)}f_{\rho_{\rm p}}(\alpha_3, \dots, \alpha_n).
\end{align}
We assume now that the remaining rapidities, $\alpha_j,~ j=3,\cdots,n$ form a set of small particle-hole excitations such that we make the following replacements 
\begin{eqnarray}
\alpha_{2j-1}\rightarrow \alpha_{j+1}+ i \pi, ~~~
\alpha_{2j} \rightarrow \alpha_{j+1} + \kappa_{j+1}, ~~~~ j=2,\cdots,m=n/2,
\end{eqnarray}
with $\kappa_{j+1}$ small. The sum over excitations can be now grouped into a sum over particle-hole pairs. For each pair, in the small excitation limit,  we have
\begin{equation}
	F(\alpha_2 | \alpha_j + \kappa_j) + F(\alpha_2|\alpha_j + i \pi )  =  F(\alpha_2|\alpha_j + \kappa_j) - F(\alpha_2 | \alpha_j) = -T^{\rm dr}(\alpha_2, \alpha_j) \kappa_j + \mathcal{O}((\kappa_j)^2),
\end{equation}
where we used that $T^{\rm dr}(\alpha_2, \alpha_j) = - \partial F(\alpha_2|\alpha_j)/\partial \alpha_j$. 
Therefore
\begin{eqnarray}
	\res{\alpha_1 = \alpha_2} f_{\rho_{\rm p}} (\alpha_1 + i \pi, \alpha_2, \alpha_3+i\pi,\alpha_3+\kappa_3, \dots, ,\alpha_{m+1}+i\pi,\alpha_{m+1}+\kappa_{m+1}) &=& \cr\cr 
 && \hskip -4.5in - \frac{\sum_{j=2}^m T^{\rm dr}(\alpha_2, \alpha_{j+1}) \kappa_{j+1} }{\rho_{\rm tot}(\alpha_1)}f_{\rho_{\rm p}}(\alpha_3+i\pi,\alpha_3, \dots,\alpha_{m+1}+i\pi,\alpha_{m+1}+\kappa_{m+1}). \label{annihilation_small_m}
\end{eqnarray}
The residue formula can be used to extract the most singular parts of a TDFF through iteration and permutation. For example, for a TDFF with $3$ particle-hole pairs, we can write
\begin{align}
    &f_{\rho_{\rm p}}(\alpha_1 + i \pi, \alpha_1 + \kappa_1, \alpha_2 + i \pi, \alpha_2 + \kappa_2, \alpha_3 + i \pi, \alpha_3 + \kappa_3) \nonumber \\
    &= \frac{V_{\rho_p}(\alpha_3)}{\rho_{\rm tot}(\alpha_1) \rho_{\rm tot}(\alpha_2)} \left(\frac{\kappa_3}{\kappa_1}T^{\rm dr}(\alpha_1, \alpha_3) + \frac{\kappa_3^2}{\kappa_1 \kappa_2} T^{\rm dr}(\alpha_1, \alpha_2)\right) T^{\rm dr}(\alpha_2, \alpha_3) + (\textrm{perm.}),
\end{align}
where $(\textrm{perm.})$ stands for summing over the permutations of indices $1,2,3$. 
We observe that it expresses the multi-particle form factor through the normalization of the single particle-hole form factors $V_{\rho_p}(\theta)$. For the operators of density and current of local conserved charges this takes a simple form, as we will discuss in Section~\ref{sec:conserved_ff}. This then provides the final answer for the singular pieces of such form factors in the small momentum limit. 

We will now turn to demonstrating that the expression~\eqref{annihilation_small_m} is consistent with what can be derived directly from the finite volume representation of the thermodynamic form factors, so validating the annihilation pole axiom. We first consider the case of two particle-hole pairs and later generalize to arbitrary number of pairs. 

\subsection{Two particle-hole pairs case}

For two particle-hole pairs, from~\eqref{annihilation_small_m}, we find
\begin{align}
	\res{\kappa_1 = 0}\, f_{\rho_{\rm p}} (\alpha_1 + i \pi, \alpha_1 + \kappa_1, \alpha_2 + i\pi, \alpha_2 + \kappa_2) =  \frac{\kappa_2 T^{\rm dr}(\alpha_1, \alpha_2)}{\rho_{\rm tot}(\alpha_1)}  f_{\rho_{\rm p}}(\alpha_2 + i\pi, \alpha_2),
\end{align}
where we have simplified $f_{\rho_{\rm p}}( \alpha_2 + i\pi , \alpha_2 + \kappa_2)$ to $f_{\rho_{\rm p}}( \alpha_2 + i\pi, \alpha_2)$ because the single particle-hole pair form factor is finite in the small excitation limit. 
We now present a derivation of this relation based on this TDFF's representation as a finite volume form factor.

We start by expressing the thermodynamic form factor through the finite volume regularization~\eqref{general_fv},
\begin{align}
	f_{\rho_{\rm p}}&(\alpha_1 + i \pi, \alpha_1 + \kappa_1, \alpha_2 + \pi i, \alpha_2 + \kappa_2) = \nonumber \\
 & = \frac{1}{2\pi \rho_{\rm tot}(\alpha_1) 2\pi \rho_{\rm tot}(\alpha_2)} \lim_{n \rightarrow \infty} \frac{f_{IV,d}(\theta_1, \dots,  \theta_n, \alpha_1, \alpha_2|  \epsilon_1, \dots, \epsilon_n, \kappa_1, \kappa_2)}{\rho_n(\theta_1, \dots, \theta_n)}+\mathcal{O}(\kappa_1)+\mathcal{O}(\kappa_2), \label{2ph_through_diagonal}
\end{align}
with  
\begin{equation} \label{epsilon_2ph}
    \epsilon_j = \frac{T^{\rm dr}(\theta_j,\alpha_1)}{L\rho_{\rm tot}(\theta_j)} \kappa_1 + \frac{T^{\rm dr}(\theta_j,\alpha_2)}{L\rho_{\rm tot}(\theta_j)} \kappa_2.
\end{equation}
We have used the definition here of the almost diagonal form factors from Section~\ref{sec:residue}. 
 In writing the above we have had to rearrange the ordering of particles.  This reordering produces corrections at  $\mathcal{O}(\kappa_1)$ and $\mathcal{O}(\kappa_2)$.  Finally we have used 
 \begin{eqnarray}
     N(\alpha + \kappa) N(\alpha + i\pi) = (2\pi L)^2 \rho_{\rm tot}^2(\alpha) + \mathcal{O}(\kappa).
 \end{eqnarray}
where this property of the normalization is developed about eq.~\ref{norm}.
We now use the residue of the almost diagonal form factors from Section 3,
\begin{align}
	&\res{\kappa_1=0}\, f_{IV,d}(\theta_1, \dots,  \theta_n, \alpha_1, \alpha_2|  \epsilon_1, \dots, \epsilon_n, \kappa_1, \kappa_2) = \nonumber \\
	  &= 2\pi \left( \kappa_2 T(\alpha_1 - \alpha_2) +  \sum_{j=1}^n \epsilon_j T(\alpha - \theta_j) \right) f_{IV,d}(\theta_1, \dots,  \theta_n, , \alpha_2|  \epsilon_1, \dots, \epsilon_n, \kappa_2).
\end{align}
The expression in the bracket, by plugging in formula~\eqref{epsilon_2ph} for $\epsilon_j$, becomes
\begin{align}
\kappa_2 T(\alpha_1 - \alpha_2) +  \sum_{j=1}^n \epsilon_j T(\alpha - \theta_j) = \kappa_2 T^{\rm dr}(\alpha_1, \alpha_2) + \mathcal{O}(\kappa_1),
\end{align}
where we ignored terms proportional to $\kappa_1$ as they do not contribute to the residue at $\kappa_1 = 0$ and took a thermodynamic limit to use the integral equation~\eqref{Tdr} for $T^{\rm}$.
The remaining almost diagonal form factor, upon dividing by $\rho_{n}(\theta_1, \dots, \theta_n)$ and in the thermodynamic limit, becomes the single particle-hole form factor. Therefore
\begin{equation}
	\res{\kappa_1=0}\, f_{\rho_{\rm p}}(\alpha_1 + i \pi, \alpha_1 + \kappa_1, \alpha_2 + \pi i, \alpha_2 + \kappa_2) = \frac{\kappa_2 T^{\rm dr}(\alpha_1, \alpha_2) }{\rho_{\rm tot}(\alpha_1)}f_{\rho_{\rm p}}(\alpha_2 + \pi i, \alpha_2 + \kappa_2),
\end{equation}
in agreement with $m=2$ case of~\eqref{annihilation_small_m}.
Repeating the computations by first taking $\kappa_2$ and then $\kappa_1$ to zero,  we would find the residue at $\kappa_2 =0$. Combining the two expressions, we capture the whole singular structure of the two particle-hole pairs form factor
\begin{align} \label{2ph_singular}
    f_{\rho_{\rm p}}&(\alpha_1 + i \pi, \alpha_1 + \kappa_1, \alpha_2 + \pi i, \alpha_2 + \kappa_2) = \nonumber \\
    & = T^{\rm dr}(\alpha_1, \alpha_2) \left( \frac{\kappa_2}{\kappa_1} \frac{f_{\rho_{\rm p}}(\alpha_2 + \pi i, \alpha_2)}{\rho_{\rm tot}(\alpha_1)} + \frac{\kappa_1}{\kappa_2} \frac{f_{\rho_{\rm p}}(\alpha_1 + \pi i, \alpha_1)}{\rho_{\rm tot}(\alpha_2)}\right) + (\dots).
\end{align}
The ellipsis $(\dots)$ represent terms of higher powers in $\kappa_1$ or $\kappa_2$. 

The expression for the singular part of the two particle-hole pair TDFF was first derived in~\cite{2018JSMTE..03.3102D} for the density operator in the Lieb-Liniger model. This is a non-relativistic theory, yet the formulas are essentially identical to the relativistic case considered here. That this structure is universal to integrable models was conjectured in~\cite{10.21468/SciPostPhys.6.4.049}. In~\cite{Cortes_Cubero_2020} this formula was derived from the axioms of the thermodynamic form factors. Here we have shown that it can also be derived from the finite volume regularization of the thermodynamic form factors.
	
\subsection{General case}

The generalization to the case of many small particle-hole excitations is straightforward. The finite volume representation gives the following expression in terms of the almost diagonal form factors
\begin{align}
	f_{\rho_{\rm p}}&(\alpha_1 + i \pi, \alpha_1 + \kappa_1, \dots, \alpha_m + i \pi, \alpha_m + \kappa_m) = \nonumber \\
 &= \prod_{j=1}^m (2\pi \rho_{\rm tot}(\alpha_j))^{-1} \lim_{\rm th} \frac{f_{IV,d}(\theta_1, \dots,  \theta_n, \alpha_1, \dots, \alpha_m|  \epsilon_1, \dots, \epsilon_n, \kappa_1, \dots, \kappa_m)}{\rho_n(\theta_1, \dots, \theta_n)},
\end{align}
with 
\begin{equation}
	\epsilon_j = \frac{1}{L  \rho_{{\rm tot}}(\theta_j)} \sum_{k=1}^m T^{\rm dr}(\theta_j, \alpha_k) \kappa_k.
\end{equation}
In the limit of small $\kappa_1$ according to the eq.~\ref{fIVd},
\begin{align}
	&\frac{f_{IV,d}(\theta_1, \dots, \theta_n, \alpha_1, \dots, \alpha_m | \epsilon_1, \dots, \epsilon_n, \kappa_1, \dots, \kappa_m)}{f_{IV,d}(\theta_1, \dots, \theta_n, \alpha_2, \dots, \alpha_m | \epsilon_1, \dots, \epsilon_n, \kappa_2, \dots, \kappa_m)} = \nonumber \\
 &= \frac{2\pi}{\kappa_1} \left( \sum_{k=2}^m \kappa_k T(\alpha_1 - \alpha_k) + \sum_{j=1}^n \epsilon_j T(\alpha_1 - \theta_j)\right) 
	= 2\pi \sum_{k=2}^m \frac{\kappa_k}{\kappa_1} T^{\rm dr}(\alpha_1, \alpha_k) + (\dots),
\end{align}
where we have neglected non-singular terms in $\kappa_1$.
This gives
\begin{align}
	&\res{\kappa_1 = 0}\,f_{\rho_{\rm p}}(\alpha_1 + i \pi, \alpha_1 + \kappa_1, \dots, \alpha_m + i \pi, \alpha_m + \kappa_m) = \nonumber \\ 
	& =\frac{1}{\rho_{\rm tot}(\alpha_1)}\left(\sum_{k=2}^m \frac{\kappa_k}{\kappa_1} T^{\rm dr}(\alpha_1, \alpha_k)\right) f_{\rho_{\rm p}}(\alpha_2 + i \pi, \alpha_2 + \kappa_2, \dots, \alpha_m + i \pi, \alpha_m + \kappa_m),
\end{align}
thus proving the small excitations case~\eqref{annihilation_small_m} of the annihilation pole axiom (the sign difference is due the residue in eq.~\eqref{annihilation_small_m} being defined as the coefficient of the $-1/\kappa_1$ term).

\section{Crossing symmetry for TDFFs}~\label{sec:crossing_small}

Until now we have considered only form factors with excitations in the {\em in} state. We introduce now the following notation for form factors with excitations in both the \emph{in}- and \emph{out}-states. For a state with 2 particle excitations in both \emph{in} and \emph{out} states, extending the notation of~\eqref{general_fv}, we write
\begin{equation}\label{gen_FF}
	f_{\rho_{\rm p}}(\beta_2, \beta_1 | \alpha_1, \alpha_2) = \lim_{td}  \frac{L^2 \langle \beta_2, \beta_1, \rho_{\rm p}| \mathcal{O}(0,0) | \rho_{\rm p}, \alpha_1, \alpha_2 \rangle}{\sqrt{\langle  \beta_2, \beta_1,\rho_{\rm p} | \rho_{\rm p}, \beta_1, \beta_2 \rangle \langle  \alpha_2, \alpha_1,\rho_{\rm p} | \rho_{\rm p}, \alpha_1, \alpha_2 \rangle}}.
\end{equation}
In the above we are considering particle excitations.  In the TDFF, a hole in the \emph{in}-state is described by a rapidity of the form $\alpha+i\pi$ while in the \emph{out}-state the rapidity denoting a hole is of the form $\beta-i\pi$.  We will see that this is consistent with the expression of the TDFFs in terms of finite volume form factors.

Crossing symmetry is a relation between two form factors in which some excitations are moved between the {\em in} and {\em out} states. For the infinite volume vacuum form factors it takes the following form~\cite{SmirnovBOOK}
\begin{eqnarray}\label{IVFFcrossing}
    f_{IV}(\beta_1, \dots, \beta_m | \alpha_1, \dots, \alpha_n) &=& f_{IV}(\alpha_1, \dots, \alpha_n, \beta_1 - i \pi, \dots, \beta_m - i \pi),\cr\cr
    f_{IV}(\beta_1, \dots, \beta_m | \alpha_1, \dots, \alpha_n) &=& f_{IV}(\beta_1 + i \pi, \dots, \beta_m + i \pi, \alpha_1, \dots, \alpha_n). 
\end{eqnarray}
The two relations are consistent with the vacuum periodicity axiom
\begin{equation}
    f_{IV}(\alpha_1, \dots, \alpha_n) = f_{IV}(\alpha_n + 2\pi i, \alpha_1, \dots, \alpha_{n-1}).
\end{equation}
Indeed, the crossing relation can be viewed as a factorization of the periodicity relation through a crossed form factor $f(\alpha_n + i \pi| \alpha_1, \dots, \alpha_{n-1})$. Our aim here is to generalize this property to thermodynamic form factors. 

For TDFFs we will show in this section that the equivalent relations to eqs.~\ref{IVFFcrossing} are
\begin{eqnarray}
f_{\rho_{\rm p}}(\beta_1, \dots, \beta_m | \alpha_1, \dots, \alpha_n) &=& f_{\rho_{\rm p}}(\beta_1 + i \pi, \dots, \beta_m + i \pi, \alpha_1, \dots, \alpha_n)\cr\cr
f_{\rho_{\rm p}}(\beta_1, \dots, \beta_m | \alpha_1, \dots, \alpha_n) &=& \prod_{i,j}R(\beta_i|\alpha_j)\prod_{i\neq j}R^{-1}(\beta_i|\beta_j)\cr\cr
&&\hskip -.5in \times f_{\rho_{\rm p}}(\alpha_1, \dots, \alpha_n, \beta_1 - i \pi, \dots, \beta_m - i \pi) .
\end{eqnarray}
The second relation directly above is a consequence of the first together with the periodicity axiom for the TDFFs.

\subsection{Crossing in TDFFs with two excitations} \label{sec:crossing_single_particles}

We begin by working out a simple example with two excitations.  We thus consider the following two form factors: $f_{\rho_{\rm p}}(\beta + i \pi, \alpha)$ and $f_{\rho_{\rm p}}(\beta| \alpha )$ with $\beta$ and $\alpha$ real. We will show that 
\begin{equation}\label{crossing_2}
    f_{\rho_{\rm p}}(\beta| \alpha ) = f_{\rho_{\rm p}}(\beta + i \pi, \alpha ).
\end{equation}
We begin by writing the expression for the finite volume regularization of a single particle-hole form factor.  We thus consider a finite volume representation of the reference state involving $n$-rapidities, $\{\theta_i\}^n_{i=1}$ where $\theta_n = \beta$, i.e., the $n$-th rapidity is where we will create the hole.  Moreover none of the $n$-rapidities are equal to $\alpha$.  Then $f_{\rho_{\rm p}}(\beta + i \pi, \alpha )$ can be defined as the following limit
\begin{eqnarray}\label{fph}
	f_{\rho_{\rm p}}(\beta + i \pi, \alpha ) &=& \lim_{td} \Gamma^{(0,0|1,1)}_n.
 \end{eqnarray}
 with $\Gamma_n$ given by 
 \begin{eqnarray}\label{Gn}
 \Gamma^{(0,0|1,1)}_n &=& \frac{L}{(N(\alpha)N^{-1}(\beta))^{1/2}}\frac{f_{IV,2n}(\beta+ \pi i, \theta_{n-1} + \pi i, \dots, \theta_1 + \pi i , \theta_1''', \dots, \theta_{n-1}''', \alpha )}{\rho_n(\theta_1, \dots, \theta_{n-1},\beta)}, 
\end{eqnarray}
with 
\begin{equation}\label{shift}
	\theta_j''' = \theta_j - \frac{F(\theta_j| \alpha)}{L \rho_{{\rm tot}}(\theta_j)} +  \frac{F(\theta_j| \beta)}{L \rho_{{\rm tot}}(\theta_j)}.
\end{equation}
and
\begin{equation}
    \rho_n(\theta_1''',\cdots,\theta_{n-1}''',\alpha) = N(\alpha)N^{-1}(\beta)\rho_n(\theta_1,\cdots,\theta_{n-1},\beta)
\end{equation}
where we are representing the state $\rho_p$ through $\{\theta\}^n_{i=1}$.

We consider now the finite volume regularization of a form factor with single particles on both sides.  Here the reference state, $\rho_p$ is given by sequences of $n$-rapidities $\{\tilde \theta_i\}^n_{i=1}$ where none of the rapidities are equal to $\alpha$ or $\beta$.  We have
\begin{equation}\label{pp1}
f_{\rho_{\rm p}}(\beta | \alpha ) = \lim_{td} \Gamma^{(1,0|1,0)}_n.
\end{equation}
with
\begin{equation}\label{pp2}
	\Gamma^{(1,0|1,0)}_n = \frac{L}{(N(\alpha)N(\beta))^{1/2}}\frac{f_{IV,2n+2}(\beta+ \pi i, \tilde\theta_n' + \pi i, \dots, \tilde\theta_1' + \pi i , \tilde\theta_1'', \dots, \tilde\theta_n'', \alpha )}{\rho_n(\tilde\theta_1, \dots, \tilde\theta_n)}, 
\end{equation}
where
\begin{equation} \label{pp3}
	\tilde\theta_j' = \tilde\theta_j - \frac{F(\tilde\theta_j| \beta)}{L \rho_{tot}(\tilde\theta_j)} , \quad \tilde\theta_j'' = \tilde\theta_j - \frac{F(\tilde\theta_j| \alpha)}{L \rho_{tot}(\tilde\theta_j)} .
\end{equation}
We now are going to show the sequences $\{\Gamma^{(1,0|1,0)}_n\}^\infty_{n=1}$ and $\{\Gamma^{(0,0|1,1)}_n\}^\infty_{n=1}$ converge to the same point.

We first note that
\begin{equation}
\rho_n(\tilde\theta_1,\cdots,\tilde\theta_n) = K^{-1}(\beta)\rho_n(\tilde\theta_1',\cdots,\tilde\theta_n') =(N(\beta))^{-1}\rho_{n+1}(\tilde\theta_1',\cdots,\tilde\theta_n',\beta) .
\end{equation}
Then $\Gamma^{(1,0|1,0)}_n$ can be written as
\begin{equation}
	\Gamma^{(1,0|1,0)}_n = \frac{L}{ (N(\alpha)N^{-1}(\beta))^{1/2}}\frac{f_{IV,2n+2}(\beta+ \pi i, \tilde\theta_n' + \pi i, \dots, \tilde\theta_1' + \pi i , \tilde\theta_1'', \dots, \tilde\theta_n'', \alpha )}{\rho_{n+1}(\tilde\theta'_1, \dots, \tilde\theta'_n,\beta)}, 
\end{equation}
If we set $\tilde\theta'_i=\theta_i, i=1,\cdots,n+1$ (which we can do -- we are free to choose $\tilde\theta_i$ in such a way so that this is true as this choice represents a valid $n+1$-rapidity representation of the reference state $\rho_p$), we see that eq.~\ref{pp3} and eq.~\ref{shift} imply that this choice leads to
\begin{equation}
    \tilde\theta'' = \theta'''.
\end{equation}
Hence with this choice $\Gamma^{(1,0|1,0)}_n=\Gamma^{(0,0|1,1)}_{n+1}$.  Thus the two limits are equal and we have established our desired crossing relation eq.~\ref{crossing_2}.

This is an expression of crossing symmetry for TDFFs, in which a particle from the {\em out} state is crossed to a hole (antiparticle) of the {\em in} state. There is a counterpart of this expression for states with single holes,
\begin{equation}  \label{crossing_1h_1h}
	f_{\rho_{\rm p}}(\alpha - i \pi | \beta + i \pi) = f_{\rho_p}(\alpha,\beta+i\pi) = S(\beta-\alpha) f_{\rho_{\rm p}}(\beta+i\pi, \alpha) .
\end{equation}
The computations are analogous to those here but for completeness, we have included the analysis in Appendix~\ref{app:extra_scattering}. To reorder the excitations in the second step we have used the scattering axiom. 

These cases demonstrate that we have consistency with how we have defined the TDFFs  in terms of their finite volume counterparts and that the hole rapidity in the \emph{out}-state should be thought of as $\alpha-i\pi$ (and not $\alpha+i\pi$). In this same Appendix~\ref{app:extra_scattering}, we also consider a form factor with a particle-hole excitation in both \emph{in}- and \emph{out}-states (so 4 excitations in total). In the following section we generalize the crossing relation to form factors with arbitrary number of excitations.

\subsection{General case}

In order to discuss the general case, we first introduce some notation.  We define indexed sets of rapidities as follows: 
 \begin{eqnarray}
      \ordered{\theta}_n = \{\theta_1, \theta_2, \dots , \theta_n\}, ~~~
       \pordered{\theta}_n = \{\theta_n, \theta_{n-1}, \dots ,\theta_1\} .
 \end{eqnarray}
In this notation, two sets that differ by a permutation of the elements are different sets. With $\pordered{\theta} = \{\theta_n, \dots, \theta_2, \theta_1\}$ we denote a set with exactly reversed order of elements to $\ordered{\theta}$. We do not assume that rapidities are ordered according to their values, $\theta_1 < \theta_2 < \theta_3 < \dots$, as is commonly done when constructing a complete basis spanned by the states $|\theta_1, \dots, \theta_n\rangle$~\cite{SmirnovBOOK,MussardoBOOK}.  

We want to establish the general crossing relation
between a form factor with $l$-particles in the \emph{out}-state and $k$-particles in the \emph{in}-state with a form factor with $k$-particles and $l$-holes in the \emph{in}-state, i.e.,
\begin{equation}\label{gen_crossing}
	f_{\rho_{\rm p}}(\orderedpi{\beta}_l, \ordered{\alpha}_k) = f_{\rho_{\rm p}}( \ordered{\beta}_l|\ordered{\alpha}_k) .
\end{equation}
To do so we first consider the TDFF on the l.h.s. of eq.~\ref{gen_crossing}.

To represent $f_{\rho_{\rm p}}(\orderedpi{\beta}_l, \ordered{\alpha}_k)$ as a limit of finite volume form factors, we consider a representation of the reference state involving $n$-rapidities, $\ordered{\theta}_n$ where $\theta_{n-i+1} = \beta_i$, i.e., the last $l$ rapidities are where we will create the $l$ holes.  Then this form factor, in direct analogy to the single particle-hole case just treated, can be defined as the following limit:
\begin{eqnarray}\label{gen_fph}
	f_{\rho_{\rm p}}(\orderedpi{\beta}_l, \ordered{\alpha}_k) &=& \lim_{td} L^{(k+l)/2}\frac{\mathcal{N}_\beta}{ \mathcal{N}_\alpha}\frac{f_{IV,2n-l+k}(\orderedpi{\beta}_l, \porderedpi{\theta}_{n-l} , \ordered{\theta'''}_{n-l}, \ordered{\alpha}_k )}{\rho_n(\ordered{\theta}_{n-l},\ordered{\beta}_l)}\cr\cr
 &\equiv& \lim_{td} \Gamma^{(0,0|k,l)}_n,
\end{eqnarray}
with the shifted $\theta'''_j$ defined as
\begin{equation}\label{gen_shift}
	\theta_j''' = \theta_j - \sum^k_{i=1}\frac{F(\theta_j| \alpha_i)}{L \rho_{{\rm tot}}(\theta_j)} +  \sum^l_{i=1}\frac{F(\theta_j| \beta_i)}{L \rho_{{\rm tot}}(\theta_j)},
\end{equation}
and the normalizations $\mathcal{N}_\alpha/\mathcal{N}_\beta$ given by
\begin{eqnarray}
   \mathcal{N}_\alpha =  
 \left(\prod_{i=1}^k N(\alpha_i)\right)^{1/2},\qquad \mathcal{N}_\beta =  
 \left(\prod_{i=1}^l N(\beta_i) \right)^{1/2}.
\end{eqnarray}
We now turn the r.h.s. of eq.~\ref{gen_crossing}.

Using an $n$-rapidity representation, $\ordered{\twt}_n$ of the reference state $\rho_p$, the finite volume regularization of the r.h.s. of eq.~\ref{gen_crossing} is
\begin{eqnarray}
f_{\rho_{\rm p}}( \ordered{\beta}_l |\ordered{\alpha}_k) &=&  \lim_{n\rightarrow \infty} L^{(k+l)/2}\frac{f_{IV,2n+k+l}(\orderedpi{\beta}_l, \porderedpi{\twt'}_{n}, \ordered{\twt''}_{n}, \ordered{\alpha}_k)}{\mathcal{N}_\alpha\mathcal{N}_\beta\rho_n(\ordered\twt_n) } \cr\cr
 &\equiv&  \lim_{td}\Gamma^{(l,0|k,0)}_n,
\end{eqnarray}
where the shifted $\twt_j$'s are defined by
\begin{equation}\label{shift2}
	\twt_j' = \twt_j - \sum_{i=1}^l\frac{F(\twt_j |\beta_i)}{L \rho_{\rm tot}(\twt_j)}, \quad \twt_j'' = \twt_j - \sum_{i=1}^k \frac{F(\twt_j | \alpha_i)}{L \rho_{\rm tot}(\twt_j)}.
\end{equation}
This is in direct analogy to eqs.~\ref{pp1}, \ref{pp2}, and \ref{pp3}.
As in the previous subsection, we reexpress the norm of the reference state as follows:
\begin{equation}
    \rho_n(\ordered{\twt}_n) = \prod^l_{i=1}K^{-1}(\beta_i)\rho_n(\ordered{\twt'}_n) = \prod^l_{i=1}(N(\beta_i))^{-1}\rho_{n+l}(\ordered{\twt'}_n,\ordered{\beta}_l) .
\end{equation}
Then $\Gamma^{(l,0|k,0)}_n$ can be written as
\begin{equation}
	\Gamma^{(l,0|k,0)}_n = L^{(k+l)/2}\frac{\mathcal{N}_\beta}{ \mathcal{N}_\alpha}\frac{f_{IV,2n+k+l}(\orderedpi{\beta}_l, \porderedpi{\twt'}_n, \ordered{\twt''}_n, \ordered{\alpha}_k)}{\rho_{n+l}(\ordered{\twt'}_n,\ordered{\beta}_l)}.
\end{equation}
As with the two-particle/1-particle-hole case, we choose $\tilde\theta_i$ so that $\tilde\theta'_i=\theta_i, i=1,\cdots,n$, note that then $\tilde\theta'' = \theta'''$ by eqs.~\ref{gen_shift} and \ref{shift2},
and so conclude $\Gamma^{(l,0|k,0)}_n=\Gamma^{(0,0|k,l)}_{n+l}$.  This gives us the general crossing relation in eq.~\ref{gen_crossing}.

\begin{figure}
    \centering
    \includegraphics[scale=0.45]{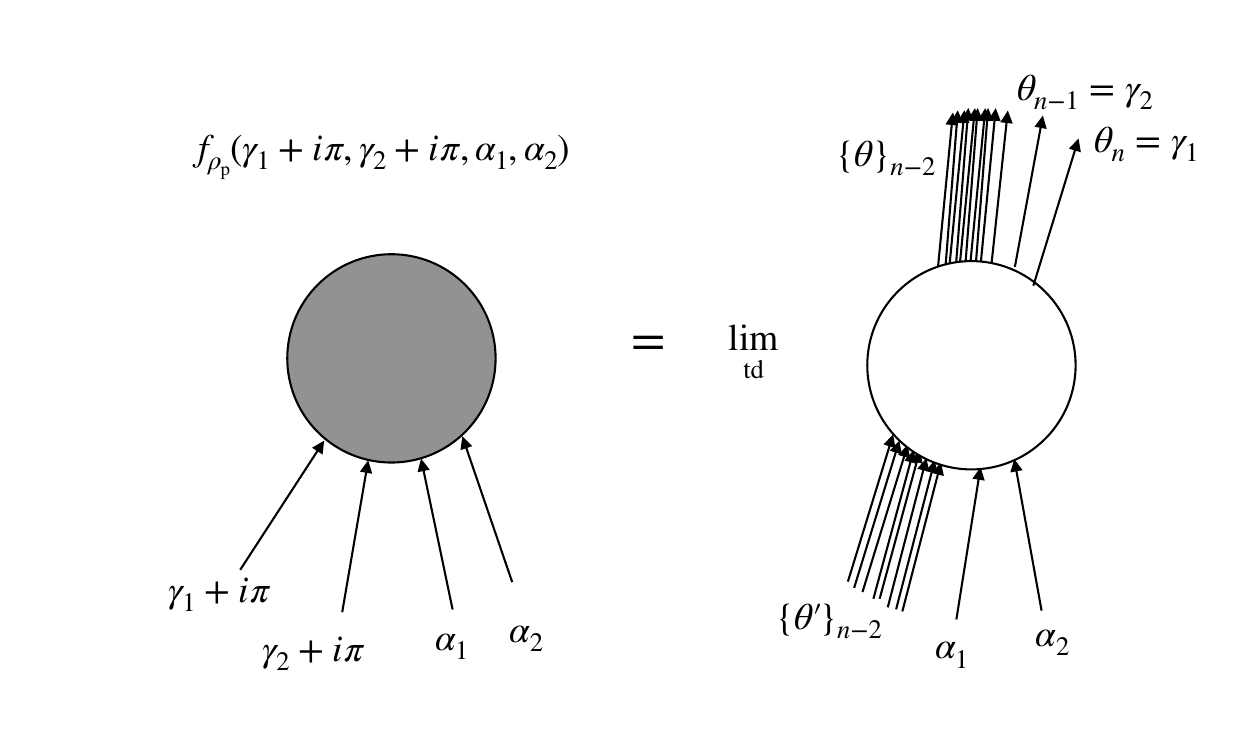}
    \caption{Visualization of the particle content of the thermodynamic form factors, here for a two particle-hole pairs excited state. The ordering of the particles follows the finite volume regularization~\eqref{general_fv}.}
    \label{fig:form_factors}
\end{figure}

\subsection{Axioms for generic form factors}
~\label{sec:crossing_axiom}

The crossing relations we have developed in the first part of this section allow us to reformulate the TDFF axioms such that they involve a generic form factor $f_{\rho_{\rm p}}(\ordered{\beta_j}|\ordered{\alpha_j})$. We find:
\begin{itemize}
	\item Scattering axiom
	\begin{align}
		f_{\rho_{\rm p}}(\beta_1,& \dots, \beta_m| \alpha_1, \dots, \alpha_j, \alpha_{j+1}, \dots, \alpha_n)  \nonumber \\
  &= S(\alpha_j - \alpha_{j+1}) f_{\rho_{\rm p}}(\beta_1, \dots, \beta_m| \alpha_1, \dots, \alpha_{j+1}, \alpha_j, \dots, \alpha_n), \\
		f_{\rho_{\rm p}}(\beta_1,& \dots, \beta_{j}, \beta_{j+1}, \dots, \beta_m| \alpha_1, \dots, \alpha_n)  \nonumber \\
  &= S(\beta_{j+1} - \beta_j) f_{\rho_{\rm p}}(\beta_1, \dots, \beta_{j+1}, \beta_j, \dots, \beta_m| \alpha_1, \dots, \alpha_n).
	\end{align}
 	\item Periodicity axiom (crossing relations)
	\begin{align}
 		f_{\rho_{\rm p}}(\beta_1&, \dots, \beta_m| \alpha_1, \dots, \alpha_n) = 
   f_{\rho_{\rm p}}(\beta_1, \dots, \beta_{m-1}| \beta_{m} + i\pi, \alpha_1, \dots, \alpha_{n-1}), \label{periodicty_axiom_1}\\
		f_{\rho_{\rm p}}(\beta_1&, \dots, \beta_m| \alpha_1, \dots, \alpha_n - i \pi)  \nonumber \\
     &= R_{\rho_{\rm p}}(\alpha_n-i\pi| \{\beta_j + i\pi\} , \{\alpha_j\}) f_{\rho_{\rm p}}(\alpha_n, \beta_1, \dots, \beta_m| \alpha_1, \dots, \alpha_{n-1}). \label{periodicty_axiom_2}
	\end{align}
        \item Annihilation pole axiom
 	\begin{eqnarray}
		- i &&\res{\beta_m \rightarrow \alpha_1 } f_{\rho_{\rm p}}(\beta_1, \dots, \beta_m| \alpha_1, \dots, \alpha_n) \cr\cr 
		&&= \frac{1}{2\pi\rho_{tot}(\beta_1)} \cr\cr 
        &&\left(1-\prod^{m-1}_{j=1}S^{dr}(\beta_j-\alpha_n)\prod^{n}_{j=2}S^{dr}(\alpha_n-\alpha_j)\right) f_{\rho_{\rm p}}(\beta_1, \dots, \beta_{m-1}| \alpha_2, \dots, \alpha_n). \label{annihilation_axiom_crossed}
	\end{eqnarray}
\end{itemize}

\begin{figure}
    \centering
    \includegraphics[scale=0.45]{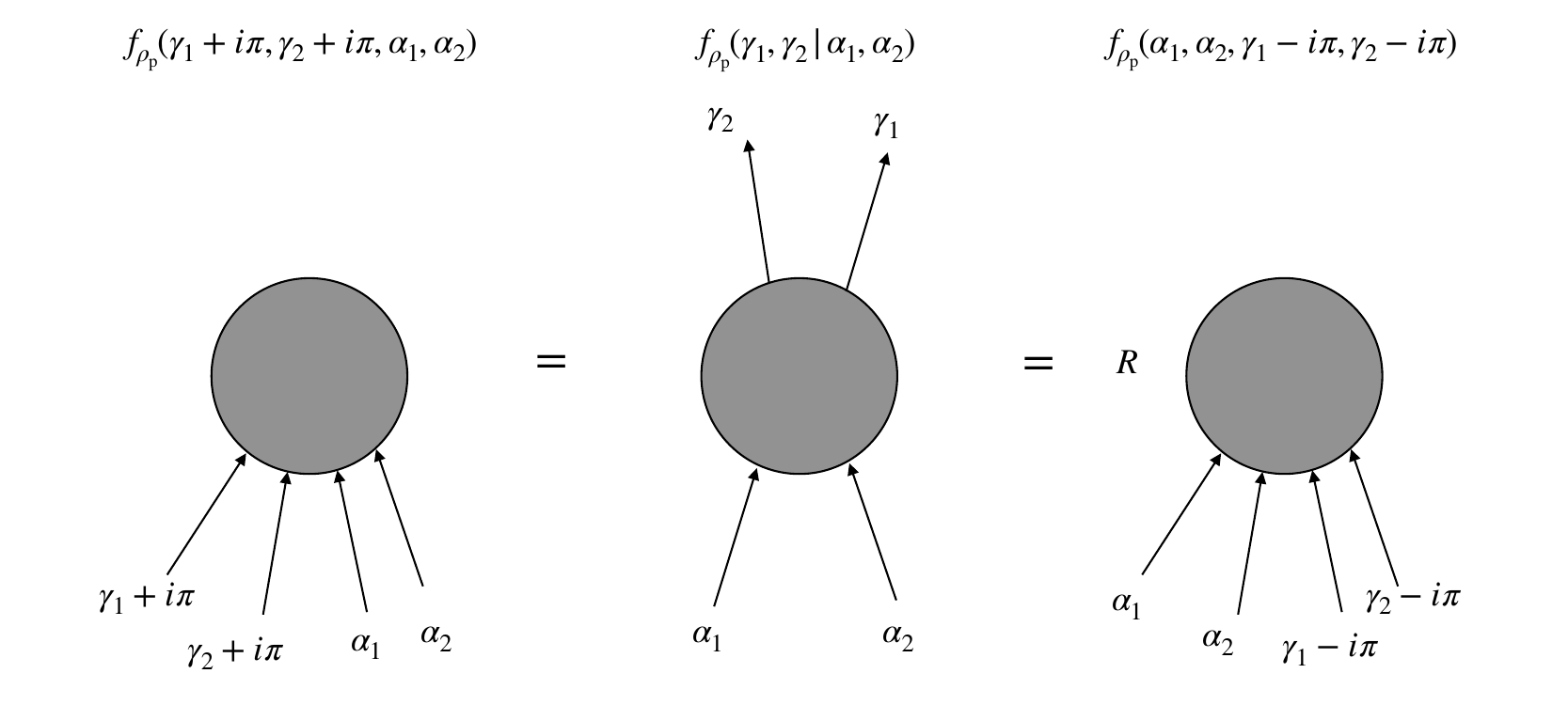}
    \caption{The crossing and periodicity relation. We visualize the three form-factors related by the crossing symmetry, eqs.~\eqref{periodicty_axiom_1} and~\eqref{periodicty_axiom_2}. The relation between the left and the right form-factors is an expression of periodicity relation which is factorized by the crossed form-factor appearing in the middle.} 
    \label{fig:crossing}
\end{figure}

\begin{figure}
    \centering
    \includegraphics[scale=0.45]{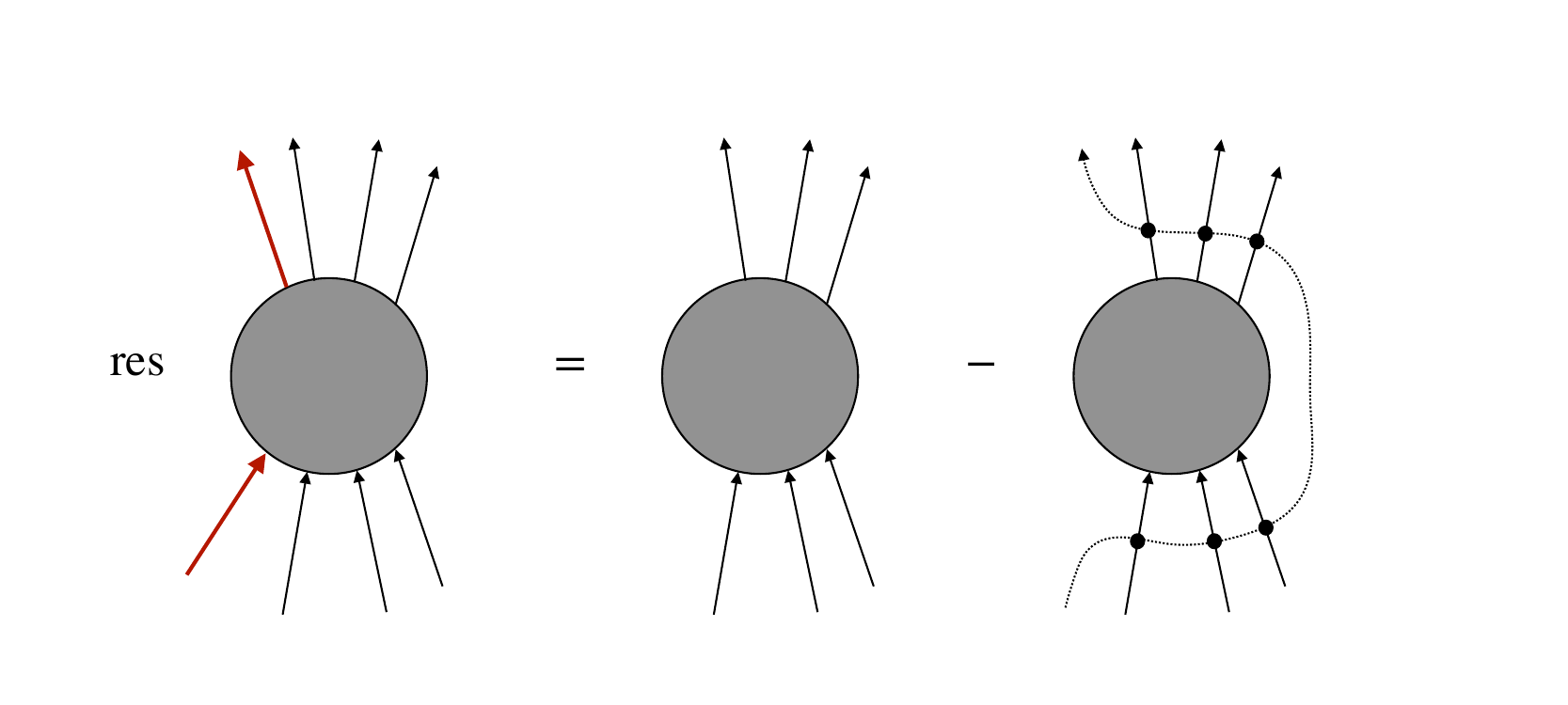}
    \caption{The annihilation axiom. The two annihilating particles are in red and the scattering (depicted by a black dot in the right form-factor) is dressed, c.f. eq.~\eqref{annihilation_axiom_crossed}.}
    \label{fig:annihilation}
\end{figure}

The axioms for the form factors, both in the vacuum and for the finite energy density states can be neatly presented visually. In fig.~\ref{fig:form_factors} we present graphical representation of the thermodynamic form factors as build up from particle excitations and background particles. The axioms for the thermodynamic form factors can be then understood through incorporating the effect of the background particles. The background particles do not have an effect on the scattering axiom, which acts \emph{locally}. On the other hand, their presence is visible in the crossing relation (fig.~\ref{fig:crossing}) and the annihilation pole axiom (fig.~\ref{fig:annihilation}) which depend on the entire particle content of the \emph{in} and \emph{out} states.

\section{Reparameterization invariance of thermodynamic form factors} \label{sec:gauge}

In this section we discuss the concept of {\em reparameterization invariance}, the independence of the TDFF of certain reparameterizations of the background state. In its full generality it can be stated through the following equality 
\begin{equation}\label{reparam}
    f_{\rho_{\rm p}}(\{\beta_j\} |  \{\alpha_j\}) 
    \equiv f_{\rho_{\rm p}}( \{\beta_j\} \cup \{\theta_i\} | \{\theta_i\} \cup \{\alpha_j\}).
\end{equation}
This equality
holds for an arbitrary {\it finite} set of $\{\theta_i\}$.  This set 
$\{\theta_i\}$ is distinct from $\{\alpha_i\}$ and $\{\beta_i\}$.  That is, we can add finite particle content as parameterized by $\{\theta_i\}$ to the thermodynamic state $\rho_{\rm p}$. Eq. \ref{reparam} defines the TDFF as equivalence classes that are infinite in extent. There is then a notion of a minimal form factor in the equivalence class.  Here minimal means that the sets of rapidities on either side of the operator in the form factor are distinct. 
In the definition of the TDFFs in eqs.~\ref{gen_FF} and \ref{general_fv}, we have added powers of $L$ that count the number of particles and holes in the form factor.  In understanding eqn.~\ref{reparam}, we {\bf do not} contribute powers of $L$ for the rapidities, here $\{\theta_i\}$, that appear identically in both the \emph{in}- and \emph{out}-states.

In adding particles to both the \emph{in}- and \emph{out}-state, reparameterization invariance follows straightforwardly from the finite volume representation of the TDFFs. As a simple example, suppose we compare $f_{\rho_p}(\beta|\alpha)$ with $f_{\rho_p}(\beta, \theta|\alpha, \theta)$, i.e. we add a single particle $\theta$ to both \emph{in}- and \emph{out}-states of $f_{\rho_p}(\beta|\alpha)$. First consider $f_{\rho_p}(\beta|\alpha)$.  Let us choose a $n$-rapidity representation, $\ordered{\theta}_n$, of the reference state $\rho_p$.  This TDFF is equal to then
\begin{eqnarray}\label{orig}
f_{\rho_p}(\beta|\alpha) &=& \lim_{td} \Gamma^{(1,0|1,0)}_n\cr\cr
\Gamma^{(1,0|1,0)}_n &=& 
L\frac{f_{2n+2,IV}(\beta+i\pi,\porderedpi{\theta'}_n,\ordered{\theta''}_n,\alpha) }{(N(\alpha) N(\beta))^{1/2}\rho_n(\ordered{\theta}_n)} .
\end{eqnarray}
with
\begin{equation}
	\theta_j' = \theta_j - \frac{F(\theta_j| \beta)}{L \rho_{tot}(\theta_j)} , \quad \theta_j'' = \theta_j - \frac{F(\theta_j| \alpha)}{L \rho_{tot}(\theta_j)} .
\end{equation}
Now consider a form-factor with a particle $\theta$ added. Describing $\rho_p$ now through a n-rapidity set $\ordered{\twt}_n$, we have
\begin{eqnarray}\label{p_add}
f_{\rho_p}(\beta, \theta|\theta, \alpha) &=& \lim_{td}\Gamma^{(2,0|2,0)}_n;\cr\cr
\Gamma^{(2,0|2,0)}_n &=& 
 L\frac{f_{2n+4,IV}(\beta+i\pi,\theta'+i\pi,\porderedpi{\twt'}_n,\ordered{\twt''}_n,\theta'',\alpha) }{N(\theta)(N(\alpha) N(\beta))^{1/2}\rho_n(\ordered{\twt})} ,
\end{eqnarray}
where we multiply $f_{2n+4,IV}$ by a single power of $L$ mindful of the rule we do not include a power of $L$ for the particle rapidity, $\theta$, as it appears identically in both the \emph{in}- and \emph{out}-state.  Here the shifts are given by
\begin{eqnarray}
	\twt_j' &=& \twt_j - \frac{F(\twt_j| \beta)}{L \rho_{tot}(\twt_j)}-\frac{F(\twt_j| \theta)}{L \rho_{tot}(\twt_j)} ;\cr\cr
 \twt_j'' &=& \twt_j - \frac{F(\twt_j| \alpha)}{L \rho_{tot}}-\frac{F(\twt_j| \theta)}{L \rho_{tot}(\twt_j)} ;\cr\cr
 \theta' &=& \theta - \frac{F(\theta| \beta)}{L \rho_{tot}(\theta)};\cr\cr
  \theta'' &=& \theta - \frac{F(\theta| \alpha)}{L \rho_{tot}(\theta)}.
\end{eqnarray}
Note that when adding particle $\theta$ we took into account shifts upon it due to particle $\alpha$ and particle $\beta$ as denoted by $\theta'$ and $\theta''$ respectively.  We do this in order to ensure that the infinite volume factor does not have any coinciding rapidities (necessitating a need to understand its disconnected pieces).  Strictly speaking we could account for shifts upon $\alpha$ and $\beta$ due to $\theta$ but because we assume that $|\alpha-\beta|\gg (mL)^{-1}$, such an accounting would only lead to subleading corrections in $1/L$.  We have also taken into account the shifts upon the rapidities describing the reference state, $\ordered{\theta}_n$, due to the addition of $\theta$.  Note however that the difference of shifts, $\twt'_j-\twt''_j$, goes unchanged from $\theta'_j-\theta''_j$.

We can think of $\Gamma^{(2,0|2,0)}_n$ in eq.~\ref{p_add} as $\Gamma^{(1,0|1,0)}_{n+1}$ in eq.~\ref{orig} if we choose the $n+1$ representation of $\rho_p$ for describing $f_{\rho_p}(\beta|\alpha)$ as follows:
\begin{equation}
    \theta_j = \twt_j - \frac{F(\twt_j|\theta)}{L\rho_{tot}(\twt_j)}; ~~~
    \theta_{n+1} = \theta.
\end{equation}
Then $\theta'_j=\twt_j'$ and $\theta''_j=\twt_j''$ and we have 
\begin{equation}
    \rho_{n+1}(\ordered{\theta}_{n+1}) = \rho_{n+1}(\{\twt_j- \frac{F(\twt_j|\theta)}{L\rho_{tot}(\twt_j)}\}^n_{j=1},\theta) =  N(\theta)\rho_n(\ordered{\twt}_n).
\end{equation}
Thus with this choice of $\ordered{\theta}_{n+1}$, we have $\Gamma^{(1,0|1,0)}_{n+1}=\Gamma^{(2,0|2,0)}_n$.  And so in the thermodynamic limit, we get as a result $f_{\rho_p}(\beta|\alpha) = f_{\rho_p}(\beta,\theta|\theta,\alpha)$.

In this demonstration that adding a particle $\theta$ to both the \emph{in}- and \emph{out}-states does not change the form-factor, we relied on $\theta$ being shifted by different amounts in the \emph{in}- and \emph{out}-states by the presence of a non-trivial particle content, i.e., $\alpha\neq\beta$.  But what happens when this shift is not present or is identical, for example, when ${\alpha_j} = {\beta_j} = \emptyset$.  Here this special case is encompassed by the LeClair-Mussardo formula.  The LeClair-Mussardo formula is by construction invariant under reparameterization:
\begin{equation}
\langle \rho_{\rm p}| \mathcal{O} | \rho_{\rm p}\rangle = f_{\rho_{\rm p}}(\{\theta_i\}| \{\theta_i\}).
\end{equation}
The presence of the extra excitations $\{\theta_j\}$ modifies the distribution $\rho_{\rm p}(\theta)$ (and the filling function $n(\theta)$) by subleading $1/L$ corrections which do not influence the evaluation of the LeClair-Mussardo series~\eqref{LcM}.

We have considered an example of reparameterization invariance where we add particles to \emph{in}- and \emph{out}-states with particle excitations.  But what happens when we add particles to \emph{in}- and \emph{out}-states containing holes and when the added particles coincide exactly with hole rapidities?  As a result we get crossing invariance.  Consider the analysis in Section 5.1 showing
\begin{equation}
    f_{\rho_p}(\beta+i\pi,\alpha)=\lim_{td} \Gamma^{(0,0|1,1)}_n;~~ f_{\rho_p}(\beta|\alpha)=\lim_{td} \Gamma^{(1,0|1,0)}_n;~~\Gamma^{(0,0|1,1)}_{n+1}=\Gamma^{(1,0|1,0)}_n.
\end{equation}
But the way we construct the TDFFs out of finite volume form factors means we can write immediately that
\begin{equation}
    f_{\rho_p}(\beta|\beta,\beta+i\pi,\alpha)=\lim_{td}\Gamma^{(1,0|1,0)}_n
\end{equation}
and crossing invariance follows.  We then see that $f_{\rho_p}(\beta|\beta,\beta+i\pi,\alpha)=f_{\rho_p}(\beta|\alpha)$ and the rapidities $\beta$ and $\beta+i\pi$ `cancel' in the \emph{in}-state.

The `cancelling' of holes by adding particles of the same rapidity to both the \emph{in}- and \emph{out}-states can be supported by the following appeal to the form factor axioms. First using crossing and then the periodicity axiom, we can write (assuming that $\alpha_1 \neq \beta_j$ for any $j$)
\begin{eqnarray}
f_{\rho_{\rm p}}(\alpha_1, \cdots , \alpha_n | \beta_1, \cdots ,\beta_k ) &=& f_{\rho_{\rm p}}(\alpha_1+i\pi, \cdots ,\alpha_n+i\pi, \beta_1, \cdots ,\beta_k ) \cr\cr 
&=& R^{-1}(\alpha_1+i\pi|\alpha_2+i\pi,\cdots,\alpha_n+i\pi,\beta_1,\cdots,\beta_k)\cr\cr  
&& \hskip .25in \times f_{\rho_{\rm p}}(\alpha_2+i\pi, \cdots ,\alpha_n+i\pi, \beta_1, \cdots ,\beta_k,\alpha_1 - i\pi),
\end{eqnarray}
where
\begin{eqnarray}\label{phase}
 R^{-1}(\alpha_1+i\pi|\alpha_2+i\pi,\cdots,\alpha_n+i\pi,\beta_1,\cdots,\beta_k) && \cr\cr 
 && \hskip -2.5in = \prod^n_{i=2}e^{-2\pi i(F(\alpha_1+i\pi|\alpha_i+i\pi)-\delta(\alpha_1-\alpha_i))}\prod^n_{i=1}e^{-2\pi i(F(\alpha_1+i\pi|\beta_i)-\delta(\alpha_1+i\pi-\beta_i))}\cr\cr 
  && \hskip -2.5in = \prod^n_{i=2}e^{-2\pi i(F(\alpha_1|\alpha_i)-\delta(\alpha_1-\alpha_i))}\prod^n_{i=1}e^{2\pi i(F(\alpha_1|\beta_i)-\delta(\alpha_1-\beta_i))},
\end{eqnarray}
where we have used the property of both the dressed and bare scattering that $F(\alpha+i\pi|\beta)=-F(\alpha|\beta)$, $\delta(\alpha+i\pi)=-\delta(\alpha)$.
If it so happens that $\alpha_i=\beta_j$ for some $i\neq 1$ and $j$, the phase factor in eq.~\ref{phase} becomes independent of $\alpha_i,\beta_j$, suggesting that the form factor itself does not depend on either of these two rapidities.

The annihilation axiom offers a similar logic:
\begin{eqnarray}
	- i \res{\beta_1 \rightarrow \beta_2 }f_{\rho_{\rm p}}(\alpha_1, \cdots , \alpha_n | \beta_1+i\pi, \beta_2,\cdots ,\beta_k ) &=&\cr\cr 
 && \hskip -2.8in \left(1 - \exp\left[2\pi i \big(\sum_{i=3}^k F(\beta_1| \beta_i) -\sum_{i=1}^k F(\beta_1| \alpha_i)\big)\right]\right)f_{\rho_{\rm p}}(\alpha_1, \cdots , \alpha_n | \beta_3, \cdots ,\beta_k ).
\end{eqnarray}
Because the contribution to the phase involves the difference of contributions coming from the $\alpha_j$'s and $\beta_i$'s, if $\beta_{i>2}=\alpha_j$ for some $i,j$, the contributions cancel and the axiom is functionally independent of $\beta_{i>2},\alpha_j$. 

One final intuitive way of understanding reparameterization invariance is to recognize that the TDFF depends on the difference between the particle content of the \emph{in} and \emph{out} states.  Thus if the \emph{in} and \emph{out}-state have coinciding content, the effects of it simply cancel.   Let us return to the equivalence we were just considering, $f_{\rho_p}(\beta|\alpha)=f_{\rho_p}(\beta+i\pi,\alpha)$.  For these two form factors, the differences between the particle distributions of the \emph{in}- and \emph{out}-states are the same.

\begin{figure}
    \centering
    \includegraphics[scale=0.3]{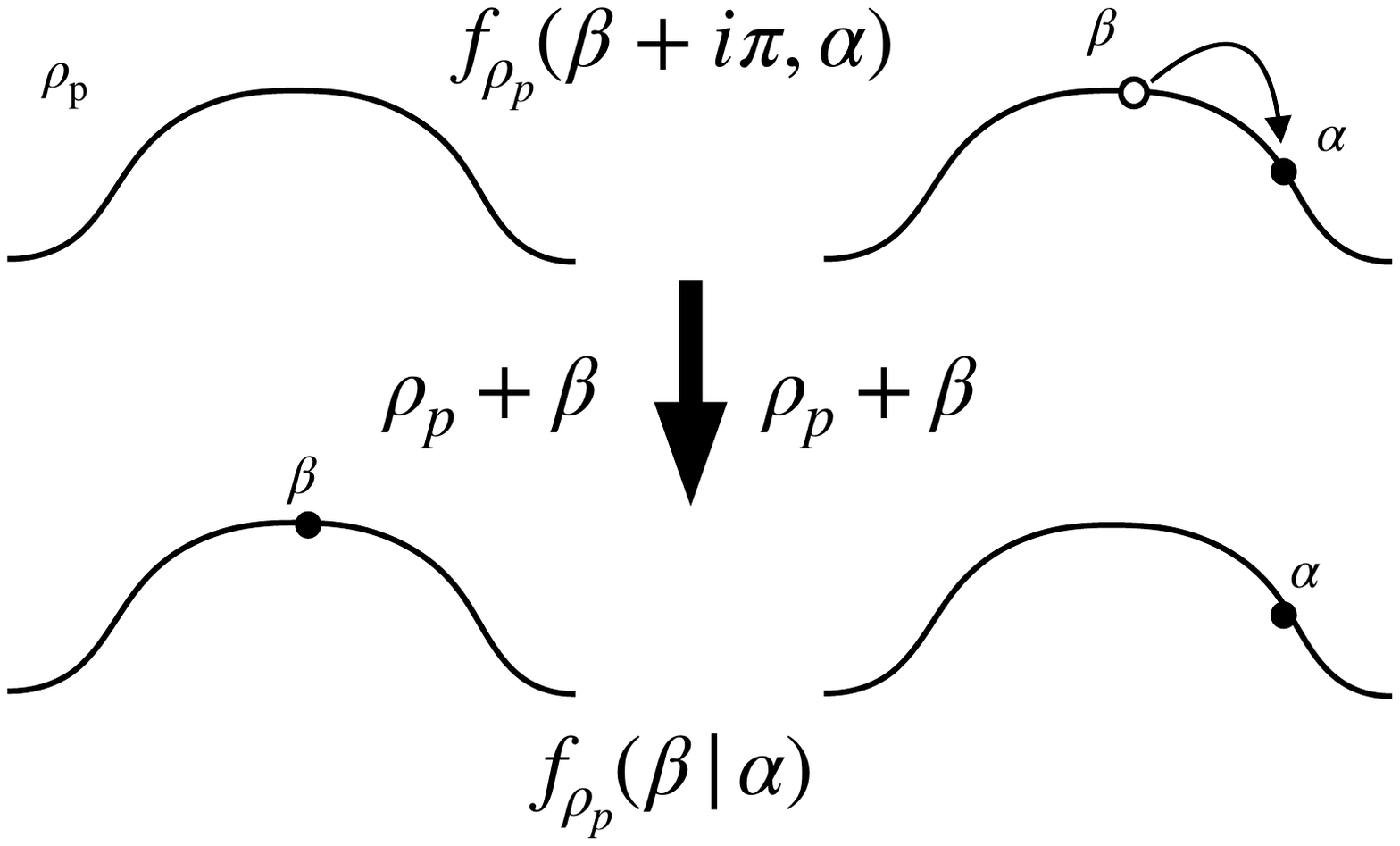}
    \caption{Illustration of how reparameterization invariance is equivalent to crossing invariance. The top part of the figure is a graphical representation of the TDFF $f_{\rho_p}(\beta-i\pi,\alpha)$ involving a particle-hole in the ket state.  We then add a particle at $\beta$ to both sides and obtain the TDFF $f_{\rho_p}(\beta|\alpha)$ where both the bra and ket states have particle and graphically represented in the bottom part of the figure.}
    \label{fig:gauge}
\end{figure}

First consider $f_\rho(\beta+i\pi,\alpha)$ with particle-hole excited state given by $(\alpha, \beta + i\pi)$. The distribution of the rapidities in the \emph{out}-state is taken to be $\rho^{out}_{\rm p}(\theta)=\rho_{\rm p}(\theta)$ while in the {\rm in}-state it is modified by $1/L$ corrections due to $\alpha$ and $\beta$:
\begin{equation}
	\rho^{in}_{\rm p}(\theta)=\rho_{\rm p}(\theta) + \frac{1}{L} \left( \delta(\alpha - \theta) - \delta(\beta - \theta)\right) + \frac{1}{L}\partial_{\theta}\left[\left( F(\theta| \alpha) - F(\theta| \beta)\right)n(\theta)\right].
\end{equation} 
We now reparametrize the TDFF by adding an excitation at $\beta$ to both the \emph{in}-state and the \emph{out}-state.  In doing so the two distributions become
\begin{eqnarray}
	\rho^{out}_{\rm p}(\theta) &=& \rho_{\rm p}(\theta) + \frac{1}{L}  \delta(\beta - \theta) + \frac{1}{L}\partial_{\theta}\left( F(\theta| \beta)n(\theta)\right), \cr\cr 
    \rho^{in}_{\rm p}(\theta) &=& \rho_{\rm p}(\theta) + \frac{1}{L} \delta(\alpha - \theta) + \frac{1}{L}\partial_{\theta}\left( F(\theta| \alpha) n(\theta)\right),
\end{eqnarray}
and both states contain now a single particle excitation, see Fig.~\ref{fig:gauge}. While $\rho^{out}_{\rm p}(\theta)$ and $\rho^{in}_{\rm p}(\theta)$ are different in both cases, their difference
\begin{equation}
\rho^{out}_{\rm p}(\theta)-\rho^{in}_{\rm p}(\theta)
\end{equation}
is the same.  

\section{Applications}

In this section we consider two applications of crossing symmetry. First we use it compute form factors of conserved charges and currents. In the second part, we use that the sinh-Gordon model in the non-relativistic limit becomes a Lieb-Liniger model to write down the crossed form factors of the particle density operator in the latter.

\subsection{Form factors of conserved densities and currents} \label{sec:conserved_ff}

The annihilation pole axiom and crossing symmetry allow us to obtain form factors with larger number of excitations from simpler ones. The situation is particularly simple for form factors of the density and current operators of conserved quantities.  Such operators typically conserve the number of particles and therefore only form factors with the same number of particles are non-zero. On the other hand the infinite series in the normalization axiom~\eqref{normalization_axiom} can be actually evaluated~\cite{2018ScPP....5...54D} providing thus a starting point from which more complicated form factors can be obtained. We discuss this now in detail.

Let us call $\hat{q}(x)$ and $\hat{j}(x)$ the conserved density and particle current respectively.  These operators satisfy a continuity equation such that
\begin{equation}
    \partial_t \hat{q}(x,t) + \partial_x \hat{j}(x,t) = 0.
\end{equation}
The action of $\hat Q = \int dx \hat q(x)$ on a state $|\theta_1,\cdots,\theta_n\rangle$ is given by
\begin{equation}
\hat{Q}|\theta_1,\cdots,\theta_n\rangle = \sum^n_{i=1} q(\theta_i)|\theta_1,\cdots,\theta_n\rangle ,
\end{equation}
where $q(\theta_i)$ is the one-particle eigenvalue of the charge operator $\hat Q$ on a particle $|\theta_i\rangle$.
The one-particle hole TDFFs $f_{\rho_{\rm p}}(\alpha+i\pi,\alpha+\kappa)$ of the density and current operators as $\kappa\rightarrow 0$ are related by
\begin{equation}
    \lim_{\kappa\rightarrow 0}f_{\rho_{\rm p}}^{\hat{j}}(\alpha+i\pi,\alpha+\kappa) = v(\alpha) \lim_{\kappa\rightarrow 0} f_{\rho_{\rm p}}^{\hat{n}}(\alpha+i\pi,\alpha+\kappa),
\end{equation}
where $v(\alpha) = \omega'(\alpha)/k'(\alpha)$. 
In the following we will consider only the density operator form factors, the current form factors can be then obtained from this relation. 
The normalization axiom~\eqref{normalization_axiom} gives~\cite{2018ScPP....5...54D}
\begin{equation} \label{normalization_applications}
    V_{\rho_{\rm p}}^{\hat{q}}(\alpha) = q_{\rm dr}(\alpha).
\end{equation}
This formula arises from an explicit summation of the LeClair-Mussardo like series appearing in the normalization axiom. For operators of local conserved charges the series has an effect of dressing the one-particle eigenvalue $q(\theta)$. For details we refer to Appendix D of~\cite{2018ScPP....5...54D}.

The crossing relations together with the small excitations limit leads then to formulas with crossed excitations. For completeness we write down all three basic form factors with two excitations:
\begin{eqnarray}
    \lim_{\kappa\rightarrow 0}f_{\rho_{\rm p}}^{\hat{q}}(\alpha+i\pi, \alpha + \kappa) &=& \lim_{\kappa\rightarrow 0}f_{\rho_{\rm p}}^{\hat{q}}(\alpha | \alpha+\kappa) = q_{\rm dr}(\alpha);\cr\cr
    \lim_{\kappa\rightarrow 0}f_{\rho_{\rm p}}^{\hat{q}}(\alpha+\kappa - i\pi | \alpha + i\pi) &=& S(0) q_{\rm dr}(\alpha).
\end{eqnarray}
As a consequence of the annihilation axiom, the form factors with two pairs of small particle-hole excitations, $(\alpha_i+\kappa_i,\alpha)$, $i=1,2$, are to leading ordering in $\kappa_i$
\begin{align}
    f_{\rho_{\rm p}}^{\hat{q}}(\alpha_2 + i \pi, \alpha_1 + \pi i, \alpha_1 + \kappa_1,  \alpha_2 + \kappa_2) =  T^{\rm dr}(\alpha_1, \alpha_2) \left( \frac{\kappa_2}{\kappa_1} \frac{q_{\rm dr}(\alpha_2)}{\rho_{\rm tot}(\alpha_1)} + \frac{\kappa_1}{\kappa_2} \frac{q_{\rm dr}(\alpha_1)}{\rho_{\rm tot}(\alpha_2)} 
    \right) + (\dots).
\end{align}
Using the crossing relation we can also write,
\begin{eqnarray}
    f_{\rho_{\rm p}}^{\hat{q}}(\alpha_1 + \kappa_1, \alpha_1 - i \pi | \alpha_2 + \pi i, \alpha_2 + \kappa_2) &&\cr\cr
    && \hskip -1.in = - T^{\rm dr}(\alpha_1, \alpha_2) \left( \frac{\kappa_2}{\kappa_1} \frac{q_{\rm dr}(\alpha_2)}{\rho_{\rm tot}(\alpha_1)} + \frac{\kappa_1}{\kappa_2} \frac{q_{\rm dr}(\alpha_1)}{\rho_{\rm tot}(\alpha_2)} 
    \right) + (\dots).
\end{eqnarray}
Here $(\dots)$ stands for terms less singular such like $\kappa_2^2/\kappa_1$ or regular in $\kappa_i$. These formulas can be also obtained directly by summing over the LeClair-Mussardo like series appearing through the finite volume regularization. 
The formulas with the crossed excitations are new.

\subsection{Form factors in the Lieb-Liniger model} \label{sec:non-relativistic}

Our work was so far focused on relativistic integrable field theories. However the thermodynamic form factors are also defined for quantum many-body systems such like the Lieb-Liniger model. This is a theory of a gas of non-relativistic bosonic particles interacting via an ultra-local potential. The Hamiltonian is
\begin{equation}
    H = \int {\rm d}x \left(- \psi^{\dagger}(x) \partial_x^2 \psi(x) + c (\psi^{\dagger}(x))^2 (\psi(x))^2 \right),
\end{equation}
with $\psi(x)$ and $\psi^{\dagger}(x)$ canonical Bose fields. Here we have taken the mass of the bosons to be $m=1/2$ and are assuming that $c>0$ where the interactions are repulsive.  An n-particle states in the Lieb-Liniger model is parameterized by momenta $\{\lambda_i\}^n_{i=1}$.  The energy and momentum of such a state are
\begin{equation}
    E = \sum^n_{i=1}\lambda_i^2, ~~~P = \sum^n_{i=1}\lambda_i.
\end{equation}
The Lieb-Liniger model can be viewed as a non-relativistic limit of the sinh-Gordon theory~\cite{2009_Kormos_PRA_81,Kormos_2010}.  The sinh-Gordon model is a model of a single real non-compact boson, $\phi(x,t)$, described by the Lagrangian,
\begin{equation}
\mathcal{L} = \frac{1}{2c_l}(\partial_t\phi)^2 - \frac{1}{2}(\partial_x\phi)^2 - \frac{m_0^2c_l^2}{g^2}\cosh(g\phi).
\end{equation}
Here $c_l$ is the speed of light, $g$ is the coupling and $m_0$ is a mass scale.  The sinh-Gordon model has a single particle which we choose to have mass $m$ (by adjusting the mass scale $m_0$ appropriately) which is its own anti-particle and has a dispersion relation described by
\begin{equation}
p(\theta) = mc_l\sinh(\theta), ~~~ E(\theta)=mc_l^2\cosh(\theta).
\end{equation}
To connect the sinh-Gordon model and the Lieb-Liniger model, one takes the double limit
\begin{equation}
    c_l\rightarrow \infty, ~~~ g\rightarrow 0, ~~~gc_l = 4\sqrt{c},
\end{equation}
subject to the constraint $gc_l = 4\sqrt{c}$.  In this limit the density of states of particles, $\rho_{tot,shG}$, in the sinh-Gordon model can be mapped onto the the density of states of the Lieb-Liniger model as follows
\begin{equation}\label{relation_rho}
    \rho_{tot,LL}(\lambda) = \widetilde\lim \frac{1}{mc_l}\rho_{tot,shG}(\theta=(mc_l)^{-1}\lambda)
\end{equation}
where $\widetilde\lim$ marks this double limit and a rapidity, $\theta$, in the sinh-Gordon theory is mapped to a rapidity, $\lambda$ in Lieb-Liniger via
\begin{equation}\label{relation_rap}
    \theta = \frac{\lambda}{mc_l}.
\end{equation}
Furthermore the shift functions of the two theories are related via
\begin{equation}\label{relation_F}
    F_{LL}(\lambda|\lambda_{\alpha})= \widetilde\lim~F_{shG}(\theta=(mc_l)^{-1}\lambda|\alpha=(mc_l)^{-1}\lambda_\alpha).
\end{equation}
The above two relations are key to connecting thermodynamic form factors in one theory to that of the other.

In this double limit normalized form factors of the Lieb-Linger density operator $\rho(x)=\psi^\dagger(x)\psi(x)$ can be written in terms of form factors of the sinh-Gordon $:\phi^2:$ operator via a simple relation
\begin{eqnarray}\label{relation_FF}
&& \frac{\langle\lambda_n,\cdots,\lambda_1|\psi^\dagger(0)\psi(0)|\lambda'_1,\cdots,\lambda'_n\rangle}{\langle\lambda_n,\cdots,\lambda_1|\lambda_1,\cdots,\lambda_n\rangle^{1/2}\langle\lambda'_n,\cdots,\lambda'_1|\lambda'_1,\cdots,\lambda'_n\rangle^{1/2}}\cr\cr
&&\hskip .025in =  \frac{1}{2}\widetilde\lim\prod^n_{j,k=1 \atop j<k}\! S(\theta_j-\theta_k)^{1/2}S(\theta'_k-\theta'_j)^{1/2} \frac{f^{:\phi^2:}_{IV}(\theta_n+i\pi,\cdots,\theta_1+i\pi,\theta'_1,\cdots,\theta'_n)}{\rho_n(\theta_1,\cdots,\theta_n)^{1/2}\rho_n(\theta'_1,\cdots,\theta'_n)^{1/2}}.
\end{eqnarray}
This formula is the normalized version of the relation between the form-factors in the two theories found in~\cite{Kormos_2010}.
The appearance of the S-matrix elements in the above reflects the fact that states in the Lieb-Liniger model are symmetrized (i.e. interchanging two $\lambda$'s leaves the state invariant), while doing the same operation for form factors in the sinh-Gordon model introduces S-matrix elements. This relation allows us to derive crossing relations for form factors in the Lieb-Liniger model in the thermodynamic limit.  

The finite system form-factors of the density operator in the Lieb-Liniger are complex numbers
\begin{equation}
    \langle\lambda_n,\cdots,\lambda_1|\psi^\dagger(0)\psi(0)|\lambda'_1,\cdots,\lambda'_n\rangle = i^n |\langle\lambda_n,\cdots,\lambda_1|\psi^\dagger(0)\psi(0)|\lambda'_1,\cdots,\lambda'_n\rangle|
\end{equation}
with the phase factor $i^n$ not having a well-defined thermodynamic limit. To have a finite thermodynamic limit we consider an absolute value of the form-factor as was done in~\cite{Smooth_us,2018JSMTE..03.3102D}. We observe that taking an absolute value of relation~\eqref{relation_FF} cancels the symmetrization factors on the sinh-Gordon side.

We thus consider a Lieb-Liniger TDFF form factor of $\psi^\dagger\psi$ defined relative to a distribution of particles $\rho_{p,LL}$
$$
f^{\psi^\dagger\psi}_{\rho_p,LL}(\lambda_{\beta_1},\cdots,\lambda_{\beta_k}|\lambda_{\alpha_1},\cdots,\lambda_{\alpha_k})
$$ 
involving $k$ particles, $\lambda_{\beta_1},\cdots,\lambda_{\beta_k}$, in the \emph{out}-state and $k$ particles, $\lambda_{\alpha_1},\cdots,\lambda_{\alpha_k}$, in the \emph{in}-state to be defined by 
\begin{eqnarray}
&& f^{\psi^\dagger\psi}_{\rho_p,LL}(\lambda_{\beta_1},\cdots,\lambda_{\beta_k}|\lambda_{\alpha_1},\cdots,\lambda_{\alpha_k}) \cr\cr
&&\hskip .25in = \lim_{td} 
 L^{k} \frac{|\langle\lambda_{\beta_1},\cdots,\lambda_{\beta_k},\lambda'_n,\cdots,\lambda'_1|\psi^\dagger(0)\psi(0)|\lambda''_1,\cdots,\lambda''_n,\lambda_{\alpha_1},\cdots,\lambda_{\alpha_k}\rangle|}
{\langle\{\lambda_{\beta_i}\},\{\lambda'_i\}|\{\lambda'_i\},\{\lambda_{\beta_i}\}\rangle^{1/2}
\langle\{\lambda_{\alpha_i}\},\{\lambda''_i\}|\{\lambda''_i\},\{\lambda_{\alpha_i}\}\rangle^{1/2}
},
\end{eqnarray}
where $(\lambda_1,\cdots,\lambda_n)$ is an $n$-particle representation of $\rho_{p,LL}(\lambda)$, $\lambda'_i,\lambda''_i$ are the shifted $\lambda$'s that result from the presence of particles in the \emph{in}- and \emph{out}-states,
\begin{equation}
    \lambda_i' = \lambda_i - \frac{1}{L\rho_{tot,LL}(\lambda_i)}\sum^k_{j=1}F_{LL}(\lambda_i|\lambda_{\beta_j}), ~~~\lambda_i'' = \lambda_i - \frac{1}{L\rho_{tot,LL}(\lambda_i)}\sum^k_{j=1}F_{LL}(\lambda_i|\lambda_{\alpha_j}),
\end{equation}
and we are using a condensed notation for the norms of the Lieb-Liniger states, i.e.,
\begin{eqnarray}
\langle\{\lambda_{\beta_i}\},\{\lambda'_i\}|\{\lambda'_i\},\{\lambda_{\beta_i}\}\rangle \equiv \langle \lambda_{\beta_1},\cdots,\lambda_{\beta_k},\lambda'_n,\cdots,\lambda'_1 | \lambda'_1,\cdots,\lambda'_n,\lambda_{\beta_k},\cdots,\lambda_{\beta_1}\rangle .
\end{eqnarray}
Now let us consider a TDFF form factor of the density operator in the Lieb-Liniger model involving $k$ particles $\{\lambda_\alpha\}$ and $k$ holes, $\{h_\beta\}$,  created at $\{\lambda_\beta\}$, all in the \emph{in}-state:
\begin{eqnarray}
&& f^{\psi^\dagger\psi}_{\rho_p,LL}(h_{\beta_1}=\lambda_{\beta_1},\cdots,h_{\beta_k}=\lambda_{\beta_k},\lambda_{\alpha_1},\cdots,\lambda_{\alpha_k}) \cr\cr
&& = \lim_{td} {L^{k}} \frac{|\langle\lambda_n=\lambda_{\beta_1},\cdots\!,\lambda_{n-k+1}=\lambda_{\beta_k},\lambda_{n-l},\cdots\!,\lambda_1|\psi^\dagger(0)\psi(0)|\lambda'''_1,\cdots\!,\lambda'''_{n-k},\lambda_{\alpha_1},\cdots\!,\lambda_{\alpha_k}\rangle|}
{\langle\{\lambda_i\}|\{\lambda_i\}\rangle^{1/2}
\langle\{\lambda_{\alpha_i}\},\{\lambda'''_i\}|\{\lambda'''_i\},\{\lambda_{\alpha_i}\}\rangle^{1/2}
},\cr&&
\end{eqnarray}
where we are choosing the finite particle representations of the reference state $\rho_{p,LL}$ to possess particles at the positions at which we create the holes (just as we do in the relativistic field theoretic case) and $\lambda_i'''$ is defined as
\begin{equation}
    \lambda_i''' = \lambda_i - \frac{1}{L\rho_{tot,LL}(\lambda_i)}\sum^k_{j=1}(F_{LL}(\lambda_i|\lambda_{\alpha_j})-F_{LL}(\lambda_i|\lambda_{\beta_j})),
\end{equation}
analogous to the definition of $\theta_i'''$ in Section 5.2.

Keeping in mind that we identify $\theta$'s in the sinh-Gordon with the $\lambda$'s in Lieb-Liniger via $\theta=\lambda/(mc_l)$ (eq. \ref{relation_rap}), we can use the mappings of the shift $F(\lambda|\alpha)$ and $\rho_{tot}$ functions between the theories (eqs. \ref{relation_F} and \ref{relation_rho}) together with the definition of the Lieb-Liniger TDFFs in terms of the sinh-Gordon TDFFs (eq. \ref{relation_FF}), to mimic the proof of the crossing relation in Section 5.2 for relativistic field theories to conclude that the Lieb-Liniger form factors also satisfy a crossing relation
\begin{eqnarray}
    f^{\psi^\dagger\psi}_{\rho_p,LL}(\lambda_{\beta_1},\cdots,\lambda_{\beta_k}|\lambda_{\alpha_1},\cdots,\lambda_{\alpha_k}) = f^{\psi^\dagger\psi}_{\rho_p,LL}(h_{\beta_1}=\lambda_{\beta_1},\cdots,h_{\beta_l}=\lambda_{\beta_k},\lambda_{\alpha_1},\cdots,\lambda_{\alpha_k}).\quad
\end{eqnarray}
This is the main result of this subsection.

\section{Conclusions}

In this work we have studied the bootstrap program for thermodynamic form factors of quantum integrable field theories. We have shown that the finite volume regularization of the form factors, which was used to formulate the normalization axiom, is consistent, with the annihilation pole axiom. This demonstrates an inner consistency of this bootstrap program. 

We have also proposed the crossing relation for the thermodynamic form factors. The proposed relation is consistent with the finite volume regularization of the thermodynamic form factors. The relation can be also argued from the \emph{reparameterization invariance}. Therefore the crossing relation can be seen as a consequence of this larger symmetry of the thermodynamic form factors. 

The crossing relation allowed us to formulate the axioms directly for a generic form factor. We used it to compute crossed form factors for the density and currents operators of local conserved charges. The resulting formulas are a new result of our work. Although the majority of our research focuses on relativistic field theories, thermodynamic form-factors can also be defined in non-relativistic systems, whether in a continuum or on a lattice.  Using the connection of the Lieb-Liniger model to the sinh-Gordon model, we have demonstrated that a crossing relation exists in this particular non-relativistic case. This finding provides additional evidence supporting the notion that the crossing relation for thermodynamic form-factors arises from an emergent reparameterization invariance, rather than relying solely on the relativistic invariance of vacuum theory. This possibility was previously suggested in~\cite{QA_Caux}.

The applications of the crossed form factors include computations of higher point correlation functions and perturbative corrections around the integrable model. These and other applications will be a subject of future work.  

The crossing relation completes the thermodynamic bootstrap program for theories with diagonal scattering and without bound states. We plan to address the form factors with bound states in the future.

\section*{Acknowledgments}
M.P. acknowledges the support from
the National Science Centre, Poland, under the SONATA
grant 2018/31/D/ST3/03588. 
R.M.K. was supported by the U.S. Department of Energy, Office of Basic Energy Sciences, under Contract No. DE-SC0012704.

\appendix

\section{Spectral representation and thermodynamic form-factors} \label{app:2pt}

In this appendix we show how the spectral representation~\eqref{2point_fnc} for the two-point function arises from a thermodynamic limit of a finite system correlation function. In finite volume the state of the system is specified by a set of quantum numbers $\{I\}$. The connected two-point function in such state is then
\begin{equation}
    \langle \{I\}|\mathcal{O}(x,t) \mathcal{O}(0) | \{I\}\rangle_{L,c} = e^{-i x P_I + i t E_I} \sum_{\{J\} \neq \{I\}} e^{i x P_J - i t E_J} |\langle \{I\} | \mathcal{O}(0) |\{J\}\rangle|^2,
\end{equation}
with the total momentum $P$ and energy $E$ expressed through the bare momenta $p(\theta)$ and energy $e(\theta)$ of eq.~\eqref{bare_kinetics}. Summation extends here over distinct sets of quantum numbers $\{J\}$ and in order to avoid the overcounting, sets differing by a permutation of their elements are not considered distinct.  
Assuming that the most important contributions to the sum come from states that not differ too much from the averaging state $| \{I\} \rangle$, we center the sum over quantum numbers $\{J\}$ around $\{I\}$. We write
\begin{equation}
    \langle \{I\} | = \langle  I_{\gamma_1},\cdots,I_{\gamma_l}, I_{n-l},\cdots,I_1|, \qquad |\{J\}\rangle = |I_1,\cdots,I_{n-l},I_{\alpha_1},\cdots,I_{\alpha_k}\rangle,
\end{equation}
with two states sharing $\{I_1, \dots, I_{n-l}\}$ quantum numbers and differing by extra quantum numbers $\{I_{\gamma_1},\cdots,I_{\gamma_l}\}$ in $\langle \{I\}|$ and $\{I_{\alpha_1},\cdots,I_{\alpha_k}\}$ in $|\{J\}\rangle$. For $k/n, l/n \ll 1$ the differences of the total momenta and energy of the two states become
\begin{align}
    P_J - P_I &= \sum_{j=1}^k k(\alpha_j) - \sum_{j=1}^l k(\gamma_j) + \mathcal{O}(1/L), \\
    E_J - E_I &= \sum_{j=1}^k \omega(\alpha_j) - \sum_{j=1}^l \omega(\gamma_j) + \mathcal{O}(1/L),
\end{align}
and involve the Dressed momenta $k(\theta)$ and energy $\omega(\theta)$. The two-point function becomes
\begin{align}
    \langle \{I\}|\mathcal{O}(x,t) \mathcal{O}(0) | \{I\}\rangle_{L,c} = \sum_{I_{\gamma} \subset I} \sum_{I_{\alpha} \subset \bar{I}} \frac{1}{|I_{\gamma}|! |I_{\alpha}|!} e^{i d(x, t; \{\alpha\}, \{\gamma\})} |\langle \{I\} | \mathcal{O}(0) |\{J\}\rangle|^2,
\end{align}
with 
\begin{equation}
    d(x,t; \{\alpha\}, \{\gamma\}) =  \sum_{j=1}^k \left(\omega(\alpha_j)t - k(\alpha_j) x\right) - \sum_{j=1}^l \left(\omega(\gamma_j)t - k(\gamma_j) x\right).
\end{equation}
The summation over $I_{\gamma}$ extends over all subsets of $I$ whereas the summations over $I_{\alpha}$ over all subsets of a set complementary to $I$, denoted $\bar{I}$. The spectral sum can be organized in terms of the cardinalities of the sets $I_\gamma$ and $I_\alpha$. The result is 
\begin{equation}
    \langle \{I\}|\mathcal{O}(x,t) \mathcal{O}(0) | \{I\}\rangle_{L,c} = \sum_{\substack{k,l=0\\ k+l>0}}^{\infty} \frac{1}{k!} \frac{1}{l!}\sum_{\substack{I_{\gamma} \subseteq I \\ |I_{\gamma}| = l}} \sum_{\substack{I_{\alpha} \subseteq \bar{I} \\  |I_{\alpha}| = k}} e^{i d(x, t; \{\alpha\}, \{\gamma\})} |\langle \{I\} | \mathcal{O}(0) |\{J\}\rangle|^2,
\end{equation}
where the condition $k+l>0$ excludes the disconneted contribution. 
In the thermodynamic limit the sum over the quantum numbers becomes an integral over the respective densities
\begin{align}
    \sum_{\substack{I_{\gamma} \subseteq I \\ |I_{\gamma}| = l}} f(I_{\gamma}) = L^l \int \prod_{j=1}^l \left( {\rm d}\gamma_j \rho_{\rm p}(\gamma_j)\right)f(\gamma_1, \dots, \gamma_l), \\
    \sum_{\substack{I_{\alpha} \subseteq \bar{I} \\ |I_{\alpha}| = k}} f(I_{\alpha}) = L^k \int \prod_{j=1}^k \left( {\rm d}\alpha_j \rho_{\rm h}(\alpha_j)\right)f(\alpha_1, \dots, \alpha_l).
\end{align}
This then leads to the following expression for the two-point function 
\begin{equation} \label{app_2pt}
    \langle \mathcal{O}(x,t) \mathcal{O}(0) \rangle_{c} = \sum_{\substack{k,l=0\\ k+l>0}}^{\infty}\fint {\rm d} \bfa_k {\rm d} \bfg_l e^{i d(x, t; \{\alpha\}, \{\gamma\})} |f_{\rho_{\rm p}}(\gamma_1+i\pi,\cdots,\gamma_l+i\pi,\alpha_1, \dots, \alpha_k)|^2,
\end{equation}
where we used the definition~\eqref{general_fv} of the thermodynamic form-factors in terms of the finite volume form-factors
\begin{eqnarray}
f_{\rho_{\rm p}}(\gamma_1+i\pi,\cdots,\gamma_l+i\pi,\alpha_1, \dots, \alpha_k) && \cr\cr
&& \hskip -2.5in = \lim_{td} L^{(k+l)/2} \langle I_{\gamma_1},\cdots,I_{\gamma_l}, I_{n-l},\cdots,I_1|\mathcal{O}|I_1,\cdots,I_{n-l} ,I_{\alpha_1},\cdots,I_{\alpha_k}\rangle.
\end{eqnarray}
In writing~\eqref{app_2pt} we introduced the integration measure
\begin{equation}
    {\rm d} \bfa_k {\rm d} \bfg_l = \frac{1}{k!} \frac{1}{l!}\prod_{j=1}^k \left(d\alpha \rho_{\rm h}(\alpha)\right) \prod_{j=1}^m \left(d\gamma_j  \rho_{\rm p}(\gamma_j)\right),
\end{equation}
and the crossed integration symbol $\fint$ refers to a Hadamard regularization procedure which is needed due to presence of the annihilation poles in the form-factors. We refer to~\cite{milosz_2021} for the details on the regularization. 

For operators conserving the number of particle only excitations with equal number of particles and holes ($k=l$) contribute to the spectral sum. In such case eq.~\eqref{app_2pt} simplifies to~\eqref{2point_fnc} presented in the introduction. 

\section{The annihilation pole from summing graphs} \label{app}

In this appendix we provide an alternative derivation of the annihilation pole axiom in a special case to that presented in Section~\ref{sec:annihilation}. We focus on the form factor with two particle-hole excitations $f_{\rho_{\rm p}}(\alpha_1 + i \pi, \alpha_1 + \kappa_1, \alpha_2 + i \pi, \alpha_2 + \kappa_2)$. We start with eq.~\eqref{2ph_through_diagonal} expressing this form factor using the finite volume regularization through an almost diagonal form factor.  To leading order in $\kappa_1,\kappa_2$ we have
\begin{eqnarray}
f_{\rho_{\rm p}}(\alpha_1 + i \pi, \alpha_1 + \kappa_1, \alpha_2 + \pi i, \alpha_2 + \kappa_2) 
&=& \frac{1}{(2\pi)^2 \rho_{\rm tot}(\alpha_1) \rho_{\rm tot}(\alpha_2)} \sum^\infty_{k=0} S_k \cr\cr
&& \hskip -3.1in  =\frac{1}{(2\pi)^2 \rho_{\rm tot}(\alpha_1) \rho_{\rm tot}(\alpha_2)} \lim_{n \rightarrow \infty} \frac{f_{IV,d}(\theta_1, \dots,  \theta_n,\theta_{n+1}= \alpha_1, \theta_{n+2}=\alpha_2|  \epsilon_1, \dots, \epsilon_n, \kappa_1, \kappa_2)}{\rho_n(\theta_1, \dots, \theta_n)}, 
\end{eqnarray}
where we define the terms $S_k$ in the above sum below. The shifts in $\theta_j$ induced by the presence of the two particle-hole pairs is given by
\begin{equation}
    \epsilon_j = \frac{T^{\rm dr}(\theta_j, \alpha_1) \kappa_1 + T^{\rm dr}(\theta_j, \alpha_2) \kappa_2}{L \rho_{{\rm tot}}(\theta_j)}.
\end{equation}
We will compute the right hand side directly using the expansion of the almost diagonal form factor into contributions from different graphs. We will do this by grouping certain graphs together and computing their thermodynamic limit. To this end we recall the formula expressing the nearly diagonal form factor in terms of graphs as discussed in Section 3.
\begin{equation}\label{f2phd}
	f_{IV,d}(\theta_1, \dots, \theta_n,\theta_{n+1}= \alpha_1, \theta_{n+2}=\alpha_2| \epsilon_1, \dots, \epsilon_n,\kappa_1,\kappa_2) = \sum_{g \in G_n} h(g).
\end{equation}
We know that there are two singular contributions to $f_{\rho_{\rm p}}(\alpha_1 + i \pi, \alpha_1 + \kappa_1, \alpha_2 + \pi i, \alpha_2 + \kappa_2)$. Here we compute only the $\kappa_1/\kappa_2$ contribution, the other contribution proportional to $\kappa_2/\kappa_1$ can be found in an analogous way.

\begin{figure}
    \centering
    \includegraphics[scale=0.6]{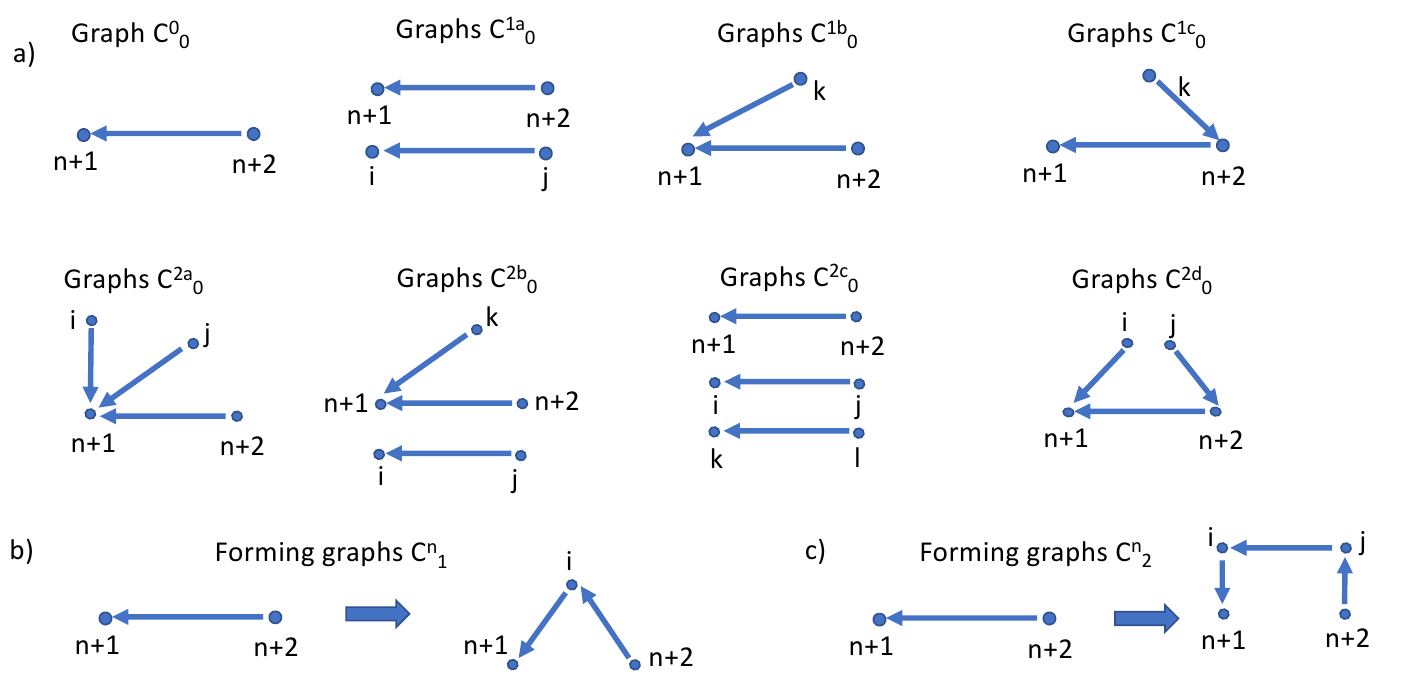}
    \caption{a) The graphs $C_0^n$ that contribute at leading order in $T(\alpha_1-\alpha_2)$. b) The substitution in a) giving graphs at next order in $T$. c) The substitution in a) giving graphs at $\mathcal{O}(T^3)$.}
    \label{fig:graphs_2ph}
\end{figure}

We now organize the contributions going as $\kappa_1/\kappa_2$ by counting the powers of $T$, the scattering kernel.  The first set, linear in $T$, is denoted by $S_0$.  The first set of diagrams, denoted by $C_0^0$, contributing to $S_0$ contains only a single graph (see Fig.~\ref{fig:graphs_2ph}) with an edge connecting $\theta_{n+2}=\alpha_2$ to $\theta_{n+1}=\alpha_1$. Its contribution to the form factor in eq.~\ref{f2phd} is
\begin{equation}
    C_0^0 = 2\pi \frac{\kappa_1}{\kappa_2} T(\alpha_1 - \alpha_2) \frac{f_c(\theta_1, \dots, \theta_n, \alpha_1,\hat{\alpha}_2)}{\rho_n(\theta_1,\cdots,\theta_n)},
\end{equation}
where $f_c(\theta_1, \dots, \theta_n, \alpha_1,\hat{\alpha}_2)$ is the connected form factor involving all of the rapidities but $\alpha_2$.
The second class of contributions, denoted $C_0^1$, contains three types of graphs. They give
\begin{align}
    C_0^1 &= C^{1a} + C^{1b} + C^{1c} = \nonumber \\
    &= 2 \pi \frac{\kappa_1}{\kappa_2} T(\alpha_1 - \alpha_2) \times 2 \pi L \sum_{i=1}^n \frac{f_c(\theta_1, \dots, \hat{\theta_i}, \dots, \theta_n, \alpha_1,\hat{\alpha}_2)\rho_{\rm tot}(\theta_i)}{\rho_n(\theta_1,\cdots,\theta_n)}.
\end{align}
Here we have used that the shifts $\epsilon_i$ obey the relation
\begin{equation}
    L \rho_{\rm tot}(\theta_i) \epsilon_i = \kappa_1 T(\theta_i - \alpha_1) + \kappa_2 T(\theta_i - \kappa_2) + \sum_{\substack{j=1, \\ j \neq i}}^n \epsilon_j T(\theta_i - \theta_j).
\end{equation}
The third class of contributions give
\begin{align}
    C_0^2 &= C^{2a} + C^{2b} + C^{2c} + C^{2d} = \nonumber \\
    &= 2 \pi \frac{\kappa_1}{\kappa_2} T(\alpha_1 - \alpha_2) \times \frac{1}{2} \sum_{i,j=1; i \neq j}^n \frac{f_c(\theta_1, \dots, \hat{\theta_i}, \dots, \hat{\theta}_j, \dots, \theta_n, \alpha_1,\hat{\alpha}_2) \tilde{\rho}_2(\theta_i, \theta_j)}{\rho_n(\theta_1,\cdots,\theta_n)},
\end{align}
where $\tilde{\rho}_2(\theta_i, \theta_j)$ is related to the finite volume norm of a two-particle state, we refer to~\cite{Cortes_Cubero_2020} for details.
The sum over the graphs $C_0^n$ can be then performed using the same techniques as in~\cite{Cortes_Cubero_2020}. The result is
\begin{eqnarray}
    S_0 &\equiv& \sum_{n=0}^{\infty} C_0^n \cr\cr
    &=& 2\pi \frac{\kappa_1}{\kappa_2} T(\alpha_1 - \alpha_2) \times \sum_{k=0}^{\infty} \frac{1}{k!}\int \prod_{j=1}^k \left(\frac{{\rm d}\theta_j}{2\pi}  n(\theta_j)\right) f_c(\theta_1, \dots, \theta_k, \alpha_1).
\end{eqnarray}
Recognising in the infinite sum the one particle hole form factor, we obtain
\begin{equation}
    S_0  = 2\pi \frac{\kappa_1}{\kappa_2} T(\alpha_1 - \alpha_2) (2\pi \rho_{tot}(\alpha_1)f_{{\rho}_{\rm p}}(\alpha_1+i\pi, \alpha_1)).
\end{equation}                                                                                                                    

This expression gives almost a correct formula for the singular part of the two particle-hole form factor. The only difference is that $T(\alpha_1 - \alpha_2)$ should be dressed. The effect of the dressing appears from considering another graphs which can be obtained from modifying those in $C_0^n$. Recall that $T^{\rm dr}$ obeys the following integral equation
\begin{equation}
    T^{\rm dr}(\alpha_1, \alpha_2) = T(\alpha_1-\alpha_2) + \int {\rm d}\mu\, T(\alpha_1 - \mu) n(\mu) T^{\rm dr}(\mu, \alpha_2).
\end{equation}
The solution can be presented in the form of the Liouville-Neumann series:
\begin{align}
        T^{\rm dr}(\alpha_1,\alpha_2) &= T(\alpha_1 - \alpha_2) + \int {\rm d}\mu\, T(\alpha_1 - \mu) n(\mu) T(\mu- \alpha_2) \nonumber \\
        & + \int {\rm d}\mu_1 {\rm d}\mu_2 T(\alpha_1 - \mu_1) n(\mu_1) T(\mu_1 - \mu_2) n(\mu_2)  T(\mu_2- \alpha_2) + \dots, 
\end{align}
 We will show now that by considering another class of diagrams, $S_1$, we find the next term of this iterative solution.

In order to do so, we consider the graphs obtained by replacing the edge $E_{n+2, n+1}$ of the graphs in Fig.~\ref{fig:graphs_2ph}a with the edges $E_{n+2,i}, E_{i, n+1}$ with $i$ being a previously edge-free vertex. This replacement is illustrated in Fig.~\ref{fig:graphs_2ph}b. We then obtain counterparts $C_1^0$, $C_1^1$, $C_1^2$ to $C_0^0$, $C_0^1$, $C_0^2$ as follows:
\begin{align}
    C_1^0 &= (2\pi)^2 \frac{\kappa_1}{\kappa_2} \left(\sum_k T(\alpha_1 - \theta_k) T(\theta_k - \alpha_2)\right) \times \frac{f_c(\theta_1, \dots, \theta_n, \alpha_1,\hat{\alpha}_2)}{\rho_n(\theta_1,\cdots,\theta_n)}, \\
    C_1^1 &= (2\pi)^2 \frac{\kappa_1}{\kappa_2} \left(\sum_k T(\alpha_1 - \theta_k) T(\theta_k - \alpha_2)\right) (2\pi L)\sum_{i=1,i\neq k}^n \frac{f_c(\theta_1, \dots, \hat{\theta_i}, \dots, \theta_n, \alpha_1,\hat{\alpha}_2)\rho_{\rm tot}(\theta_i)}{\rho_n(\theta_1,\cdots,\theta_n)} \\
    C_1^2 &= (2\pi)^2 \frac{\kappa_1}{\kappa_2} \left(\sum_k T(\alpha_1 - \theta_k) T(\theta_k - \alpha_2)\right)\\
    &\hskip 1in \times \frac{1}{2} \sum_{i,j=1; i \neq j;i,j\neq k}^n \frac{f_c(\theta_1, \dots, \hat{\theta_i}, \dots, \hat{\theta}_j, \dots, \theta_n, \alpha_1,\hat{\alpha}_2) \tilde{\rho}_2(\theta_i, \theta_j)}{\rho_n(\theta_1,\cdots,\theta_n)}\nonumber.
\end{align}
If we sum up $C_1^n$ we obtain
\begin{eqnarray}
    S_1 &\equiv& \sum_{n=0}^{\infty} C_1^n = 2\pi \frac{\kappa_1}{\kappa_2} \sum_k \frac{1}{L\rho_{tot}(\theta_k)}T(\alpha_1-\theta_k)T(\theta_k-\alpha_2)\cr\cr
    &&\hskip 1in \times (2\pi \rho_{tot}(\alpha_1)f_{{\rho}_{\rm p}}(\alpha_1+i\pi, \alpha_1)) ,
\end{eqnarray}
which indeed contains the discretized first order correction to the iterative solution of $T^{\rm dr}$.

If we now sum $S_0, S_1, S_2, \dots$ we obtain higher order corrections and ultimately $T^{\rm dr}$:
\begin{equation}
    f_{\rho_{\rm p}}(\alpha_1 + i \pi, \alpha_1 + \kappa_1, \alpha_2 + \pi i, \alpha_2 + \kappa_2) = \frac{1}{\rho_{tot}(\alpha_2)}\frac{\kappa_1}{\kappa_2} T^{\rm dr}(\alpha_1 - \alpha_2) f_{{\rho}_{\rm p}}(\alpha_1+i\pi, \alpha_1),
\end{equation}
in agreement with~\eqref{2ph_singular}.

\section{Crossing symmetry for simple excited states} \label{app:extra_scattering}

In this Appendix we present complementary computations to those presented in Section~\ref{sec:crossing_single_particles}. We consider form factors with a hole excitation in both states and with particle-hole excitation in both states. We again establish the crossing relations here using the finite-volume representation of the TDFFs. 

\subsection{Hole excitations on both sides}

We consider a TDFF with holes on either side of the operator, i.e., $f_{\rho_p}(\alpha-i\pi|\beta+i\pi)$.  Note that the hole in the \emph{out}-state at $\alpha$ is given by $\alpha-i\pi$ in the TDFF whereas the hole in the {\emph in}-state at $\beta$ is given by $\beta+i\pi$.  To describe this TDFF in terms of a finite volume form factor, we choose an n-rapidity representation of $\rho_p$ such that
\begin{equation}
    \twt_1,\cdots,\tilde\theta_{n-2},\tilde\theta_{n-1}=\alpha,\tilde\theta_n=\beta,
\end{equation}
i.e., the state has rapidities at both $\alpha$ and $\beta$ so that holes can be in fact created.
The finite volume regularization is then
\begin{eqnarray}\label{fhh}
    f_{\rho_p}(\alpha-i\pi|\beta+i\pi) &=& \lim_{td} S(\beta-\alpha)  \Gamma^{(0,1|0,1)}_n;\cr\cr
    \Gamma^{(0,1|0,1)}_n &=&  \frac{L f_{IV,2n-2}(\beta +  i \pi, \twt_{n-2}' + \pi i, \dots, \twt_1' + \pi i , \twt_1'', \dots, \twt_{n-2}'', \alpha)}{\rho_n(\twt_1, \dots, \twt_{n-2},\alpha,\beta)(N(\alpha)N(\beta))^{-1/2}},
\end{eqnarray}
where $\twt'_j,\twt''_j$ are defined by
\begin{equation}\label{deftwtprime}
	\twt_j' = \twt_j + \frac{F(\twt_j| \alpha)}{L \rho_{tot}(\twt_j)}, \quad \twt_j'' = \twt_j + \frac{F(\twt_j| \beta)}{L \rho_{tot}(\twt_j)}  .
\end{equation}
We note that 
\begin{equation}
    \rho_n(\twt_1, \dots, \twt_{n-2},\alpha,\beta)= K(\alpha)\rho_n(\twt'_1, \dots, \twt'_{n-2},\alpha,\beta),
\end{equation}
which allows us to write $\Gamma^{(0,1|0,1)}_n$ as 
\begin{equation}
    \Gamma^{(0,1|0,1)}_n =  \frac{L(2\pi L\rho_{tot}(\alpha)N(\beta))^{1/2}}{K^{1/2}(\alpha)} \frac{f_{IV,2n-2}(\beta +  i \pi, \twt_{n-2}' + \pi i, \dots, \twt_1' + \pi i , \twt_1'', \dots, \twt_{n-2}'', \alpha)}{\rho_n(\twt'_1, \dots, \twt'_{n-2},\alpha,\beta)}.
\end{equation}
The prefactor of $S(\beta-\alpha)$ appears in eq.~\ref{fhh} because in order to create the hole at $\alpha$ in the \emph{out}-state, we need to reorder the rapidities $\theta_n=\beta$ and $\theta_{n-1}=\alpha$ via the scattering relation.

Now let us consider the crossed version of this form factor, i.e., $f_{\rho_{\rm p}}(\beta + i \pi, \alpha )$.  From eq.~\ref{fph} and eq.~\ref{Gn} we have that
\begin{eqnarray}
	f_{\rho_{\rm p}}(\beta + i \pi, \alpha ) &=& \lim_{td} \Gamma^{(0,0|1,1)}_n.
 \end{eqnarray}
 with $\Gamma^{(0,0|1,1)}_n$ given by 
 \begin{eqnarray}
 \Gamma^{(0,0|1,1)}_n &=& \frac{L}{(N(\alpha)N^{-1}(\beta))^{1/2}}\frac{f_{IV,2n}(\beta+ \pi i, \theta_{n-1} + \pi i, \dots, \theta_1 + \pi i , \theta_1''', \dots, \theta_{n-1}''', \alpha )}{\rho_n(\theta_1, \dots, \theta_{n-1},\beta)}, 
\end{eqnarray}
where the $\theta_j'''$'s are defined in eq.~\ref{shift}.
Here we can add a particle $\alpha$ to $\rho_n$ obtaining
\begin{equation}
    \rho_n(\theta_1, \dots, \theta_{n-1},\beta) = (2\pi L\rho_{tot}(\alpha))^{-1}\rho_{n+1}(\theta_1, \dots, \theta_{n-1},\alpha,\beta)
\end{equation}
This allows us to write $\Gamma^{(0,0|1,1)}_n$ as
\begin{eqnarray}
 \Gamma^{(0,0|1,1)}_n &=& \frac{L(2\pi L\rho_{tot}(\alpha)N(\beta))^{1/2}}{K^{1/2}(\alpha)}\frac{f_{IV,2n}(\beta+ \pi i, \theta_{n-1} + \pi i, \dots, \theta_1 + \pi i , \theta_1''', \dots, \theta_{n-1}''', \alpha )}{\rho_n(\theta_1, \dots, \theta_{n-1},\alpha,\beta)}. \cr\cr &&
\end{eqnarray}
We now use our freedom to choose $\twt_j$ so that $\twt'_j=\theta_j$.  By eq.~\ref{deftwtprime} and eq.~\ref{shift}, we see then that $\twt''_j=\theta'''_j$.  Hence $\Gamma^{{(0,1|0,1)}}_{n+1}=\Gamma^{(0,0|1,1)}_n$ and so we have
\begin{equation}
    f_{\rho_p}(\alpha-i\pi|\beta+i\pi)=S(\beta-\alpha)f_{\rho_p}(\beta+i\pi,\alpha).
\end{equation}
We note that this could have been derived by analytically continuing the crossing relation derived in Section 5.2,
\begin{equation}
    f_{\rho_p}(\alpha|\beta)=f_{\rho_p}(\alpha+i\pi,\beta),
\end{equation}
and using the scattering axiom.  Our definition of the TDFFs in terms of finite volume form factors thus passes an important consistency check.

\subsection{TDFFs with particle-hole excitations on both sides of the operator}

Here we will establish the crossing relation for a TDFF with a particle-hole pair in both the \emph{in}- and \emph{out}-states. In particular, we will show that
\begin{equation}
    f_{\rho_p}(\alpha_1,\beta_1-i\pi|\beta_2+i\pi,\alpha_2) = f_{\rho_p}(\alpha_1+i\pi,\beta_1,\beta_2+i\pi,\alpha_2),
\end{equation}
by again appealing to the representation of the TDFFs as limits of finite volume form factors.

To represent $f_{\rho_p}(\alpha_1,\beta_1-i\pi|\beta_2+i\pi,\alpha_2)$ as a finite volume form factor, we choose an $n$-rapidity state where we can create holes at $\beta_1$ and $\beta_2$:
\begin{equation}
    \theta_1,\cdots,\theta_{n-2},\theta_{n-1}=\beta_1,\theta_{n}=\beta_2.
\end{equation}
With such a representation, we can write $f_{\rho_p}(\alpha_1,\beta_1-i\pi|\beta_2+i\pi,\alpha_2)$ as 
\begin{eqnarray}\label{f2ph}
    f_{\rho_p}(\alpha_1,\beta_1-i\pi|\beta_2+i\pi,\alpha_2) &=& S(\beta_2-\beta_1)\lim_{td} \Gamma^{(1,1|1,1)}_n;\cr\cr
    && \hskip -2.4in \Gamma^{(1,1|1,1)}_n = L^2\bigg(\frac{N(\beta_1)N(\beta_2)}{N(\alpha_1)N(\alpha_2)}\bigg)^{1/2}\cr\cr
    && \hskip -1.5in \times \frac{f_{IV,2n}(\alpha_1+i\pi,\beta_2 + i\pi, \theta'_{n-2} + \pi i, \dots, \theta'_1 + i\pi , \theta_1'', \dots, \theta_{n-2}'', \beta_1,\alpha_2 )}{\rho_n(\theta_1, \dots, \theta_{n-2},\beta_1,\beta_2)},
\end{eqnarray}
where $\theta'_j,\theta''_j$ are defined by
\begin{eqnarray}
	\theta_j' &=& \theta_j - \frac{F(\theta_j| \alpha_1)}{L \rho_{tot}(\theta_j)} + \frac{F(\theta_j| \beta_1)}{L \rho_{tot}(\theta_j)} ;\cr\cr
 \theta_j'' &=& \theta_j - \frac{F(\theta_j| \alpha_2)}{L \rho_{tot}(\theta_j)} + \frac{F(\theta_j| \beta_2)}{L \rho_{tot}(\theta_j)} .
\end{eqnarray}
The prefactor of $S(\beta_2-\beta_1)$ appears in eq.~\ref{f2ph} because again we need to commute the position of $\beta_1$ and $\beta_2$ in order to create the hole in the \emph{out}-state.

We can rewrite $\rho_n$ in the above in terms of a state norm with two more particles (at $\alpha_1$ and $\alpha_2$)
\begin{equation}
    \rho_n(\theta_1, \dots, \theta_{n-2},\beta_1,\beta_2)= \frac{K(\beta_1)}{(N(\alpha_1)2\pi L\rho_{tot}(\alpha_2)}\rho_{n+2}(\theta'_1, \dots, \theta'_{n-2},\beta_1,\beta_2,\alpha_1,\alpha_2),
\end{equation}
which then allows us in turn to rewrite $\Gamma^{(1,1|1,1)}_n$ as 
\begin{eqnarray}
    \Gamma^{(1,1|1,1)}_n =  L^2\bigg(\frac{ N(\alpha_1)2\pi L\rho_{tot}(\alpha_2) 2\pi L\rho_{tot}(\beta_1)N(\beta_2)}{K(\alpha_2)K(\beta_1)}\bigg)^{1/2} \cr\cr
    && \hskip -3.5in \times \frac{f_{IV,2n}(\alpha_1+i\pi,\beta_2 + i\pi, \theta'_{n-2} + \pi i, \dots, \theta'_1 + i\pi , \theta_1'', \dots, \theta_{n-1}'', \beta_1,\alpha_2 )}{\rho_{n+2}(\theta_1', \dots, \theta_{n-2}',\beta_1,\beta_2,\alpha_1,\alpha_2)}.
\end{eqnarray}
We now turn to considering the TDFF with both particle-hole pairs in the \emph{in}-state, i.e., $f_{\rho_p}(\alpha_1+i\pi,\beta_2+i\pi,\beta_1,\alpha_2)$.  

To do so, we chose an n-rapidity representation of $\rho_p$ so that there are two rapidities, $\beta_2$ and $\alpha_1$, where we want to create the holes in the crossed version of the TDFF:
\begin{equation}
    \twt_1,\cdots,\twt_{n-2},\twt_{n-1}=\beta_2,\theta_{n}=\alpha_1.
\end{equation}
With this choice we have
\begin{eqnarray}
    f_{\rho_p}(\alpha_1+i\pi,\beta_2+i\pi,\beta_1,\alpha_2) &=& \lim_{td} \Gamma^{(0,0|2,2)}_n;\cr\cr
    && \hskip -2.4in \Gamma^{(0,0|2,2)}_n = L^2\bigg(\frac{N(\alpha_1)N(\beta_2)}{N(\beta_1)N(\alpha_2)}\bigg)^{1/2}\cr\cr
    && \hskip -1.5in \times \frac{f_{IV,2n}(\alpha_1+i\pi,\beta_2 + i\pi, \twt_{n-2} + i \pi, \dots, \twt_1 + i\pi , \twt_1''', \dots, \twt_{n-2}''', \beta_1,\alpha_2 )}{\rho_n(\twt_1, \dots, \twt_{n-2},\beta_2,\alpha_1)}.
\end{eqnarray}
Here $\twt'''_j$ is given by
\begin{equation}
    \twt'''_j = \twt +\frac{1}{L\rho_{tot}(\twt_j)}(F(\twt_j|\alpha_1)+F(\twt_j|\beta_2)-F(\twt_j|\alpha_2)-F(\twt_j|\beta_1)).
\end{equation}
Thus $\twt'''_j-\twt_j=\theta_j''-\theta_j'$.

 We now rewrite the denominator of $\Gamma^{(0,0|2,2)}_n$ as
 \begin{equation}
     \rho_n(\twt_1, \dots, \twt_{n-2},\beta_2,\alpha_1)= \frac{\rho_{n+2}(\twt_1, \dots, \twt_{n-2},\beta_2,\alpha_1,\beta_1,\alpha_2)}{4\pi^2 L^2\rho_{tot}(\alpha_2)\rho_{tot}(\beta_1)}.
 \end{equation}
This allows us to express $\Gamma^{(0,0|2,2)}_n$ as
\begin{eqnarray}
    \Gamma^{(0,0|2,2)}_n = L^2\bigg(\frac{4\pi^2L^2\rho_{tot}(\alpha_2)\rho_{tot}(\beta_1) N(\alpha_1)N(\beta_2)}{K(\beta_1)K(\alpha_2)}\bigg)^{1/2}\cr\cr
    && \hskip -3.0in \times \frac{f_{IV,2n}(\alpha_1+i\pi,\beta_2 + i\pi, \twt_{n-2} + i \pi, \dots, \twt_1 + i\pi , \twt_1''', \dots, \twt_{n-2}''', \beta_1,\alpha_2 )}{\rho_{n+2}(\twt_1, \dots, \twt_{n-2},\beta_2,\alpha_1,\beta_1,\alpha_2)}.
\end{eqnarray}
Because $\twt'''_j-\twt_j=\theta_j''-\theta_j'$, we can choose the $n$-rapidity representatives of $\rho_p$ so that $\theta'_j=\twt_j$ and conclude that
\begin{equation}
    \Gamma^{(0,0|2,2)}_n = \Gamma^{(1,1|1,1)}_n.
\end{equation}
Hence
\begin{equation}
    f_{\rho_p}(\alpha_1+i\pi,\beta_2+i\pi,\beta_1,\alpha_2)= S(\beta_1-\beta_2)f_{\rho_p}(\alpha_1,\beta_1-i\pi|\beta_2+i\pi,\alpha_2).
\end{equation}
Using the scattering axiom on the l.h.s. of the above, we obtain the crossing relation in its final form:
\begin{equation}
    f_{\rho_p}(\alpha_1+i\pi,\beta_1,\beta_2+i\pi,\alpha_2)= f_{\rho_p}(\alpha_1,\beta_1-i\pi|\beta_2+i\pi,\alpha_2).
\end{equation}
This is consistent with the general crossing relation in eq.~\ref{gen_crossing}.

\section{CCD factors}

The axioms for the thermodynamic form-factors do not specify them uniquely. It is easily seen that if $f_{\rho_p}(\alpha_1, \dots, \alpha_n)$ satisfies the axiom then
\begin{equation}
    f_{\rho p}(\alpha_1, \dots, \alpha_n) W(\alpha_1, \dots, \alpha_n)
\end{equation}
also satisfies them for any symmetric function $W(\alpha_1, \dots, \alpha_n)$ such that 
\begin{equation} \label{CCD_conditions}
    \lim_{\alpha_1 \rightarrow \alpha_2 + i \pi} W(\alpha_1, \dots, \alpha_n) = W(\alpha_3, \dots, \alpha_n), \qquad \lim_{\alpha_1 \rightarrow \alpha_2 + i \pi} W(\alpha_1, \alpha_2) = 1.
\end{equation}

In the computations~\cite{Smooth_us,Bootstrap_JHEP} of the thermodynamic form-factors, it was suggested that their definition involves a resummation over the so-called soft-modes. In the simplest situation this resummation leads to a multiplicative factor 
\begin{equation} \label{diff_entropy_factor}
    W(\alpha_1, \dots, \alpha_n) = \exp\left( \sum_{i=1}^n \delta S(\alpha_i)\right),
\end{equation}
where $\delta S(\alpha_i)$ is the differential entropy~\cite{KorepinBOOK},
\begin{equation}
    \delta S(\alpha) = \int {\rm d}\theta s(\theta) \frac{\partial}{\partial \theta} \left( \frac{F(\theta|\alpha}{\rho_{\rm tot}(\theta)}\right).
\end{equation}
The differential entropy obeys $\delta S(\alpha + i\pi) = - \delta S(\alpha)$. Therefore $W(\alpha_1, \dots, \alpha_n)$ obeys the conditions~\eqref{CCD_conditions} and is a symmetric function. This shows that the entropical factor stemming from the resummation procedure modifies the form-factors in a way compatible with the axioms.

\bibliographystyle{JHEP}
\bibliography{bib}

\end{document}